\documentclass[12pt]{elsart}

\usepackage{amsfonts}

\usepackage{psfig}

\def\bfk{\mbox{\mathversion{bold}${k}$}}
\def\bfm{\mbox{\mathversion{bold}${m}$}}

\def\bfkm{{\mbox{\mathversion{bold}${k}$}_{\mbox{\mathversion{bold}$\scriptstyle{m}$}}}}

\def\eikdotx{e^{i \mbox{\mathversion{bold}$\scriptstyle{k\cdot x}$}}}
\def\eikjdotx{e^{i \mbox{\mathversion{bold}$\scriptstyle{k}$}_j\cdot 
                   \mbox{\mathversion{bold}$\scriptstyle{x}$}}}
\def\eikmdotx{e^{i \mbox{\mathversion{bold}$\scriptstyle{k}$}_{\mbox{\mathversion{bold}$\scriptstyle{m}$}}\cdot 
                   \mbox{\mathversion{bold}$\scriptstyle{x}$}}}
\def\eix{e^{ix}}
\def\emix{e^{-ix}}
\def\enix#1{e^{#1ix}}
\def\emnix#1{e^{-#1ix}}
\def\calFmu{{\mathcal F}_\mu}
\def\calN{{\mathcal N}}
\def\calNmu{{\mathcal N}_\mu}
\def\calNn{{\mathcal N}_n}
\def\calLmu{{\mathcal L}_\mu}
\def\calLz{{\mathcal L}_0}
\def\calLzinv{{\mathcal L}_0^{-1}}
          
\newcommand{\semidirectprodsymbol}{ %
       \mathbin{\vrule height1.10ex depth 0.05ex width0.25pt
       \mkern-2.9mu \mathchar"0202}}    
 
\newcommand{\sdp}{ \mskip-\medmuskip \mkern5mu
       \mathbin{\semidirectprodsymbol} \penalty 900 \mkern5mu \mskip-\medmuskip}

\begin{document}
\begin{frontmatter}
\centerline{For submission to Physica D\hfill}

\title{Convergence properties of the 8, 10 and 12 mode representations of
quasipatterns}

\author{A.M.~Rucklidge}
\address{Department of Applied Mathematics,
University of Leeds, Leeds LS2 9JT, UK}
\and
\author{W.J.~Rucklidge}
\address{148 Promethean Way, Mountain View, CA 94043, USA}

 \begin{abstract}
Spatial Fourier transforms of quasipatterns observed in Faraday wave
experiments suggest that the patterns are well represented by the sum of 8, 10
or 12 Fourier modes with wavevectors equally spaced around a circle. This
representation has been used many times as the starting point for standard
perturbative methods of computing the weakly nonlinear dependence of the
pattern amplitude on parameters. We show that nonlinear interactions of
$n$~such Fourier modes generate new modes with wavevectors that approach the
original circle no faster than a constant times~$n^{-2}$, and that there are
combinations of modes that do achieve this limit. As in KAM theory, small
divisors cause difficulties in the perturbation theory, and the convergence of
the standard method is questionable in spite of the bound on the small
divisors. We compute steady quasipattern solutions of the cubic
Swift--Hohenberg equation up to $33^{\mbox{$\scriptstyle{\mathrm{rd}}$}}$~order
to illustrate the issues in some detail, and argue that the standard method
does not converge sufficiently rapidly to be regarded as a reliable way of
calculating properties of quasipatterns.
 \end{abstract}


\end{frontmatter}

\section{Introduction}

Experimental observations of regular patterns have been widely reported in many
physical systems, for example, Rayleigh--B\'enard convection,
reaction--diffusion problems and the Faraday wave
experiment~\cite{refC119,refR70}. In the last example, a tray containing a
layer of fluid is subjected to vertical vibrations, and the flat horizontal
surface of the fluid becomes unstable once the amplitude of the vibration
exceeds a critical value. With multi-frequency forcing, this experiment is
capable of producing a wide variety of patterns with an astonishing degree of
symmetry~\cite{refA65}.

The simplest patterns: stripes, squares and hexagons, have reflection, rotation
and translation symmetries, and a comprehensive theory has been developed to
analyse the creation of these patterns from the initial flat
state~\cite{refG59}. In order to apply the theory to the experiments, two
idealisations are necessary: first, the experimental boundaries are ignored,
and so in effect the experiment is supposed to be taking place in a container
of infinite size: there are two unbounded spatial directions; and second, the
observed pattern is supposed to have perfect spatial periodicity. Restricting
to a spatially periodic subdomain enables rigorous theory to be
applied~\cite{refC23}, and the existence of stripe, square, and hexagon (and
other) solutions of model partial differential equations (PDEs) can be proven
using equivariant bifurcation theory~\cite{refG59}. Given that in some highly 
controlled experiments the idealisation of spatial periodicity appears to 
hold over dozens of wavelengths of the pattern, these assumptions are perfectly 
reasonable when the objective is to understand the nature of these periodic
patterns.

 \begin{figure}
 \begin{center}
 \mbox{\makebox[0.25\hsize][c]{(a)}
       \hspace{0.05\hsize}
       \makebox[0.25\hsize][c]{(b)}
       \hspace{0.05\hsize}
       \makebox[0.25\hsize][c]{(c)}}
 \mbox{\psfig{file=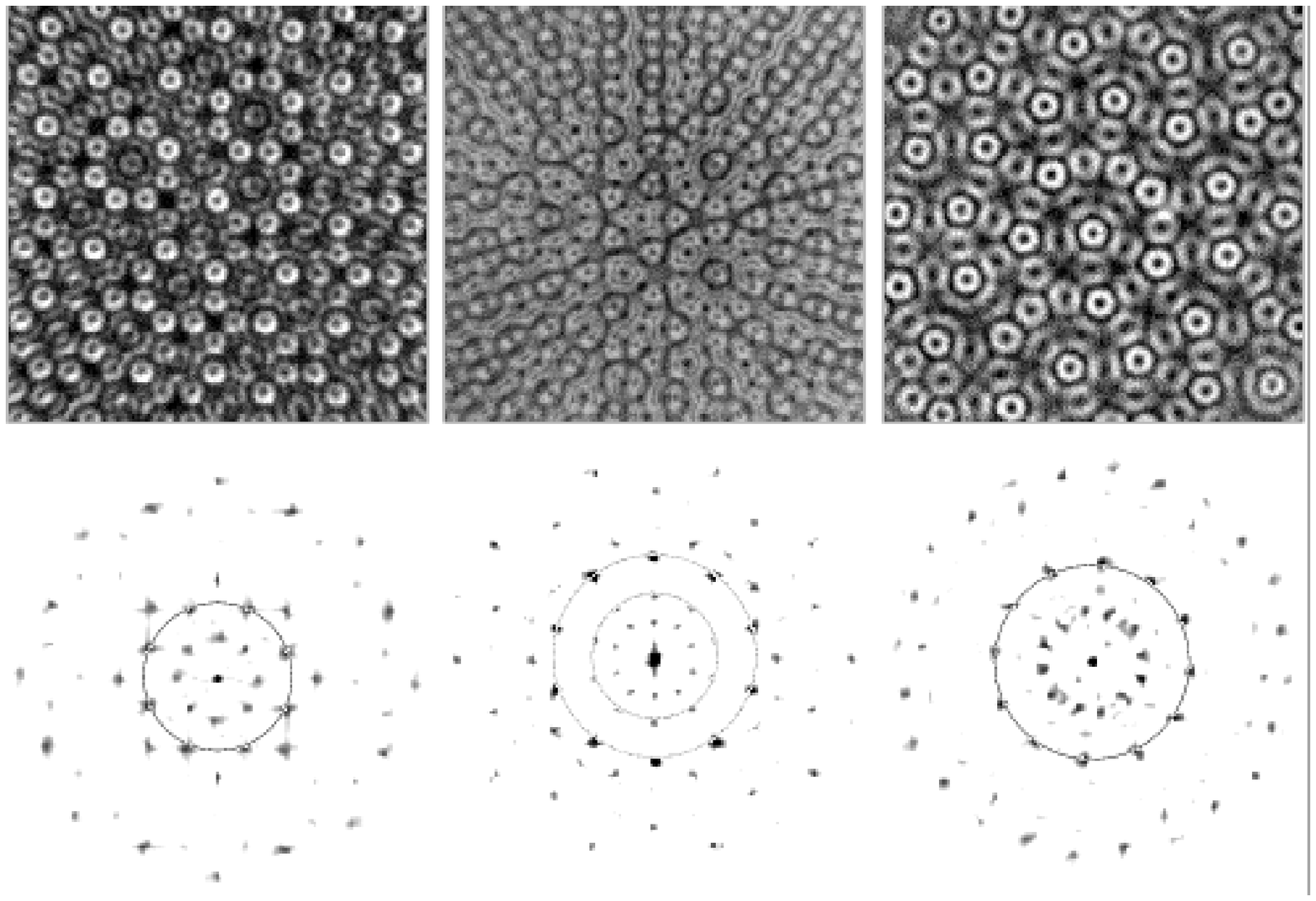,width=0.9\hsize}}
 \mbox{\makebox[0.25\hsize][c]{(d)}
       \hspace{0.05\hsize}
       \makebox[0.25\hsize][c]{(e)}
       \hspace{0.05\hsize}
       \makebox[0.25\hsize][c]{(f)}}
 \mbox{\psfig{file=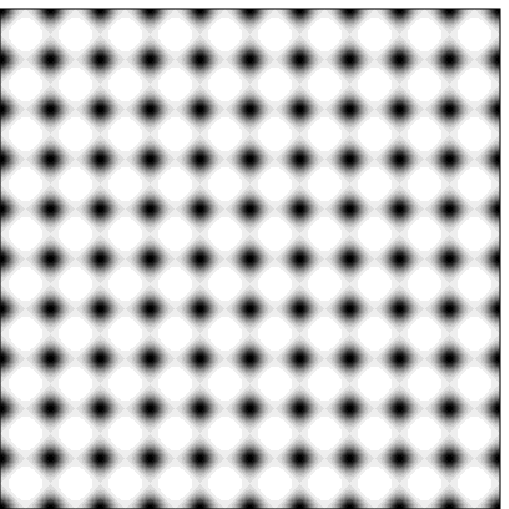,width=0.25\hsize}
       \hspace{0.05\hsize}
       \psfig{file=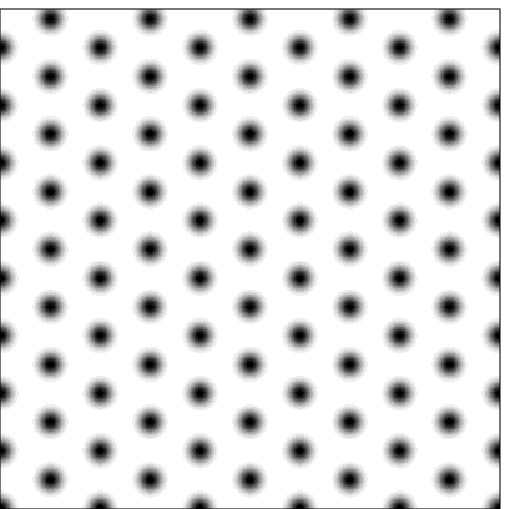,width=0.25\hsize}
       \hspace{0.05\hsize}
       \psfig{file=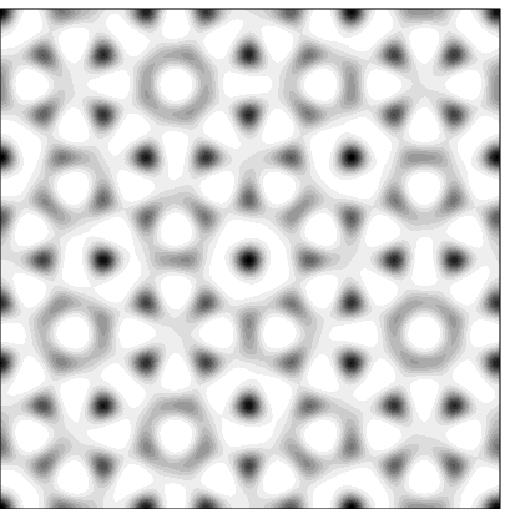,width=0.25\hsize}}
 \mbox{\makebox[0.25\hsize][c]{(g)}
       \hspace{0.05\hsize}
       \makebox[0.25\hsize][c]{(h)}
       \hspace{0.05\hsize}
       \makebox[0.25\hsize][c]{(i)}}
 \mbox{\psfig{file=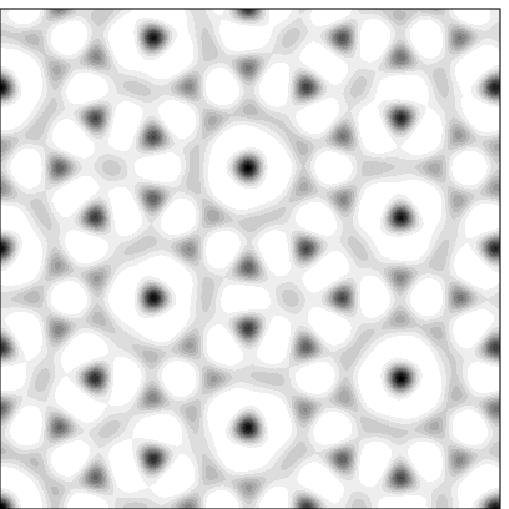,width=0.25\hsize}
       \hspace{0.05\hsize}
       \psfig{file=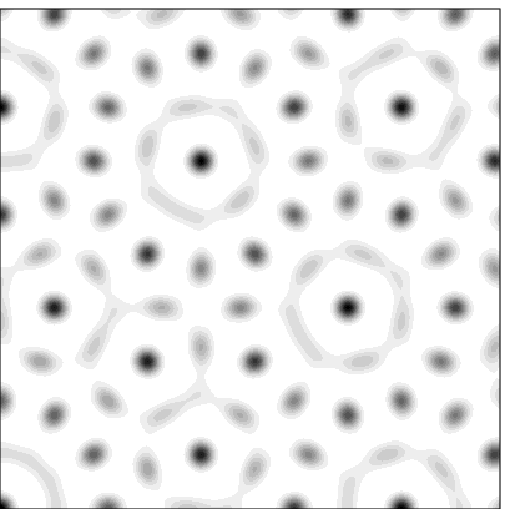,width=0.25\hsize}
       \hspace{0.05\hsize}
       \psfig{file=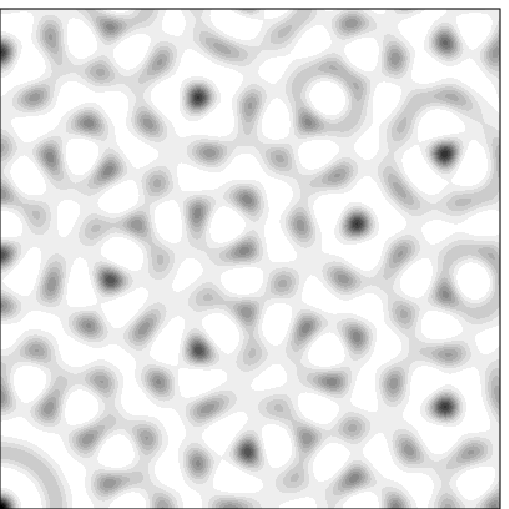,width=0.25\hsize}}
 \end{center}
 \caption{(a--c)~Experimental observation of (a)~8-fold, (b)~10-fold, and
 (c)~12-fold quasipatterns in the
 Faraday wave experiment with (a)~3 and (b,c)~2 frequency forcing. 
 Top line: experimental photographs; bottom line: spatial Fourier transform.
 The circles in the Fourier spectra indicate wavenumbers that are excited by 
 one of the harmonics in the forcing (both harmonics in case~b).
 From Arbell and Fineberg (2002)~\cite{refA65}, with permission. 
 (d--i)~Synthetic examples of periodic patterns (d,e) with $Q=4$, 6 
 wavevectors with equal amplitudes, and quasipatterns (f,g,h,i) with
 $Q=8$, 10, 12, 14 wavevectors.}
 \label{fig:qpexamples}
 \end{figure}

However, experiments are quite capable of producing patterns that cannot be
analysed in this way. Notable examples of this include {\em quasipatterns},
which are quasiperiodic in any spatial direction, that is, the amplitude of the
pattern (taken along any direction in the plane) can be regarded as a sum of
waves with incommensurate spatial frequencies. Experimental photographs of
quasipatterns are reproduced in figure~\ref{fig:qpexamples}(a--c)
from~\cite{refA65}; the lack of spatial periodicity can be seen in the images,
and the long range rotational order is evident from the spatial Fourier
transforms, in a manner similar to quasicrystals~\cite{refJ41}. The Faraday
wave experiment has been a particularly fruitful source of quasipatterns since
they were discovered by Christiansen \etal~\cite{refC118} (8-fold quasipattern
with single frequency forcing and a relatively low viscosity fluid) and Edwards
and Fauve~\cite{refE15,refE10} (12-fold quasipattern with two frequency forcing
and a high viscosity fluid). See references~\cite{refA65,refK67,refB90,refB91}
for further large aspect ratio experiments, and~\cite{refM118} for a recent
review of experimental and theoretical issues. Quasipatterns have also been
reported in studies of nonlinear optical systems~\cite{refP54,refL65}, and in
numerical studies of several model PDEs~\cite{refZ9,refM126,refL59}, though of
course computations carried out in large periodic domains can only approximate
a quasiperiodic pattern.

The forcing in Faraday wave experiments can be a simple sinusoid, but
quasipatterns are more readily generated when two or three commensurate
temporal frequencies are included in the forcing. Each frequency excites, or
nearly excites, waves with a particular wavenumber, and nonlinear resonant
interactions between these waves, as well as waves that damped, can be used to
encourage or discourage particular waves to appear in the
pattern~\cite{refE10,refS105,refT60,refP53}. By tuning such parameters as the
driving frequency, the fluid viscosity and the layer depth, experimentalists
have been able to produce very clean examples of quasipatterns (as in
figure~\ref{fig:qpexamples}a--c), as well as to demonstrate a clear
understanding of the physical mechanisms behind their production.

Since quasipatterns, by their very nature, do not fit into periodic domains,
equivariant bifurcation theory cannot be applied, and other methods are
required in order to predict, for example, the dependence of the amplitude of
the quasipattern on parameters, or the stability of the quasipattern. One
approach, which has been followed many times, is to suppose that the dynamics
of the quasipattern is dominated by the evolution of the amplitudes of
$Q$~waves ($Q$~even), with wavevectors distributed equally around a circle.
Equations governing the evolution of these amplitudes can readily be written
down, and take the form
 \begin{equation}\label{eq:naiveamplitude}
 {\dot A_j} = \mu A_j + \sum_{k=1}^{Q/2} \beta_{j,k}\left|A_k\right|^2A_j
                      + \hbox{resonant terms},
 \end{equation}
where $A_j$~is the complex amplitude of mode~$j$, $\mu$~describes the forcing
of the pattern, and the $\beta_{j,k}$~coefficients depend on the angle between
the wavevectors of modes $j$ and~$k$. Resonant terms (depending on the value
of~$Q$) are also included but not written explicitly above. 
Steady solutions of~(\ref{eq:naiveamplitude}) with all amplitudes equal 
represent patterns of the type shown in figure~\ref{fig:qpexamples}(d--i), for  
$Q=4$, \dots, 14 wavevectors.

Equation~(\ref{eq:naiveamplitude}) can be written down directly from symmetry
or general physical considerations~\cite{refM130,refN18,refS110,refE16}, but it
has also been derived from PDEs that model the hydrodynamic and other
problems~\cite{refP56,refG95,refL66,refC120}. The method that is used is called
modified perturbation theory~\cite{refM25,refS63}, and the amplitude
equation~(\ref{eq:naiveamplitude}) is only the leading order approximation to
the equations that govern the evolution of the amplitudes of the modes.

The problem with this approach when applied to quasipatterns is that it
overlooks the near-singular resonances that small divisors can cause, and
ignores the nearly neutral modes that are driven by high-order nonlinear
interactions.

The difficulty of small divisors arises in a variety of situations as well as 
this one, for example, the persistence of quasiperiodic oscillations in
Hamiltonian systems, the analysis of which cumulated in the KAM theorem
(cf.~\cite{refM125}). To take an illustrative example, consider the
one-dimensional ordinary differential equation with quasiperiodic forcing:
 \begin{equation}
 \frac{da}{dt} = \sum_{m_1,m_2=-\infty}^{\infty}
                 C_{m_1,m_2}e^{i(m_1\omega_1+m_2\omega_2)t},
 \end{equation}
where $C_{m_1,m_2}$ are constants satisfying~$C_{0,0}=0$ and $C_{m_1,m_2}\leq
K_1\left(|m_1|+|m_2|\right)^{-\gamma}$ for every pair of integers~$m_1$
and~$m_2$, with $\gamma>2$ so that the sum converges. The frequencies
$\omega_1$ and $\omega_2$ are incommensurate. This series can be integrated
formally term by term to give:
 \begin{equation}
 a(t) = \sum_{m_1,m_2=-\infty}^{\infty}
                          \frac{C_{m_1,m_2}}{i(m_1\omega_1+m_2\omega_2)}
                          e^{i(m_1\omega_1+m_2\omega_2)t}.
 \end{equation}
It is clear that this sum for~$a(t)$ may not converge even if the sum for the
forcing function does, since $m_1\omega_1+m_2\omega_2$ comes arbitrarily close
to zero, and so the amplitudes of the Fourier coefficients can be arbitrarily
large. However, if $\omega_1$ and $\omega_2$ satisfy a Diophantine condition,
that is, if there are constants~$K_2>0$ and $\delta>0$ such that $\omega_1$ and
$\omega_2$ satisfy
 \begin{equation}
 \left|m_1\omega_1+m_2\omega_2\right| \geq 
                                      K_2\left(|m_1|+|m_2|\right)^{-\delta}
 \end{equation}
for every $m_1$ and~$m_2$, then the sum for~$a(t)$ can readily be shown to
converge provided~$\gamma>2+\delta$. This increases the constraints
on the smoothness of the forcing function. Normally, integrating a Fourier
series poses no difficulties, but this example demonstrates that when
quasiperiodic functions are involved, an extra degree of caution is necessary.

Currently, there is no KAM-like theory for quasipatterns, which are
quasiperiodic in two dimensions~$(x,y)$, rather than than the usual one
dimension (time). There is, however, a theory for one-dimensional steady
quasipatterns~\cite{refI7}, which makes use of space as a time-like evolution
variable. In the absence of a rigorous theory for two-dimensional
quasipatterns, we examine the issue of convergence or otherwise of the
standard method of computing amplitude equations of the form
of~(\ref{eq:naiveamplitude}), when applied to quasipatterns.

In section~\ref{sec:PerturbationTheory}, we revisit the standard method of
modified perturbation theory as applied to the computation of the amplitude of
a quasipattern as a function of a parameter close to the onset of the pattern,
and point out where the problem of small divisors arises. In
section~\ref{sec:Modes}, we work out just how small the small divisors are
(with numerical results that make use of a rapid method presented in
Appendix~\ref{app:RapidMethod}), and in section~\ref{sec:Convergence} return to
general issue of convergence. We discuss the specific example of the
Swift--Hohenberg equation in section~\ref{sec:SwiftHohenberg}, and conclude in
section~\ref{sec:Discussion} with the argument that the standard method does
not appear to converge sufficiently rapidly to be regarded as a reliable way of
calculating properties of quasipatterns.

\section{Perturbation theory}
\label{sec:PerturbationTheory}

In order to point out exactly where the difficulties lie, we begin by going
through the standard modified perturbation theory~\cite{refM25,refS63} 
for a general pattern-forming PDE:
 \begin{equation}
 \frac{\partial U}{\partial t} = \calFmu(U) 
                               = \calLmu(U) + \calNmu(U), 
 \end{equation}
where $U(x,y,t)$ represents the order parameter (or any measure of the
pattern), $\calFmu$ is an operator containing spatial derivatives that
depends on a parameter~$\mu$ and that can be split into linear 
($\calLmu$) and nonlinear ($\calNmu$) parts. 
The order parameter may be 
multi-dimensional. A specific example is the Swift--Hohenberg
equation~\cite{refS109}:
 \begin{equation}\label{eq:SwiftHohenberg}
 \frac{\partial U}{\partial t} = \mu U - (1+\nabla^2)^2 U - U^3,
 \end{equation}
where $U(x,y,t)\in\Rset$, but many pattern-forming problems can be cast into
this form, or variations~\cite{refM99}. 

The spatially uniform trivial state $U(x,y,t)=0$ is always a possible solution,
but it loses stability as the parameter~$\mu$ increases through a critical
value~$\mu_0$, which we take to be zero. We focus on the case where the mode
that becomes unstable is a Fourier mode with nonzero wavenumber, which we scale
to~1. The problem is posed on the whole plane, so $(x,y)\in\Rset^2$, but if we
were interested only in spatially periodic solutions, then the whole plane
could be restricted to a periodic domain, and standard equivariant bifurcation
theory~\cite{refG59} could be used. However, this rules out the spatially
quasiperiodic patterns of interest here.

Instead of equivariant bifurcation theory, we use the older technique of
modified perturbation theory~\cite{refM25,refS63}, and suppose that the
parameter~$\mu$ is close to its critical value~$\mu_0=0$, that the amplitude of
the solution is small and that the pattern evolves slowly. We introduce a small
parameter~$\epsilon$, scale time by~$\epsilon^{-1}$, and write
 \begin{equation}\label{eq:powerseriesU}
 U = \epsilon U_1 + \epsilon^2 U_2 + \epsilon^3 U_3 + \ldots,\qquad
 \mu = \epsilon \mu_1.
 \end{equation}
Then $\calLmu(U)$ is of order~$\epsilon$, and $\partial U/\partial T$
and $\calNmu(U)$ are of order~$\epsilon^2$. In many examples, there
are additional symmetries in the problem, and it may be necessary to scale time
by~$\epsilon^{-2}$ and set $\mu=\epsilon^2\mu_2$ if it turns out that
$\mu_1=0$. We focus on the general case here, treating the specific example of
the Swift--Hohenberg equation in section~\ref{sec:SwiftHohenberg}.

\begin{figure}
 \begin{center}
 \mbox{\makebox[2.0truein][c]{(a)}
       \hspace{1.0truein}
       \makebox[2.0truein][c]{(b)}}
 \mbox{\psfig{file=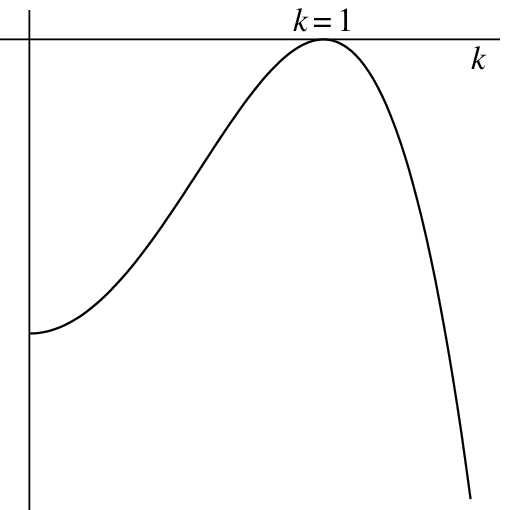,width=2.0truein}
       \hspace{1.0truein}
       \psfig{file=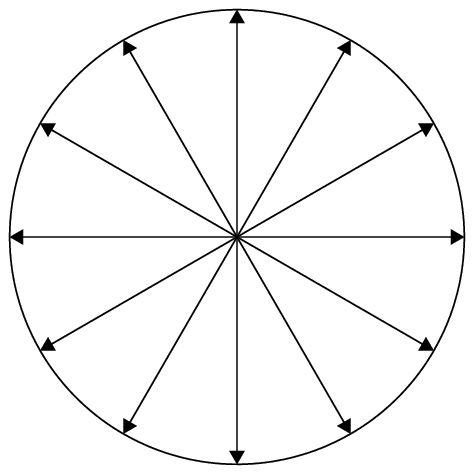,width=2.0truein}}
 \end{center}
 \caption{(a)~Schematic growth (decay) rate of a mode $\eikdotx$, as a function
of~$k=|\bfk|$ at $\mu=0$. Modes with $k=1$ are marginally stable. (b)~$Q=12$
wavevectors on the circle~$k=1$.}
 \label{fig:stabilitylattice}
 \end{figure}

The leading order equation, at order~$\epsilon$, is
 \begin{equation}\label{eq:epsi1}
 \calLz(U_1)=0.
 \end{equation}
In the Swift--Hohenberg example, the operator $\calLz$ would 
be~$-(1+\nabla^2)^2$. Since $\mu=0$ is a bifurcation point, the linear
operator $\calLz$ is singular with a circle of marginally stable 
Fourier modes in its kernel: $\calLz(\eikdotx)=0$ whenever
$k=|\bfk|=1$, (see figure~\ref{fig:stabilitylattice}a) and so~(\ref{eq:epsi1}) 
has nontrivial solutions of the form
 \begin{equation}
 U_1(x,y,t)=\sum_{j=1}^Q A_j(t)\eikjdotx,
 \end{equation}
where we can select any $Q$~wavevectors~$\bfk_j$, with $j=1,\dots,Q$, from the
circle~$k=1$ (figure~\ref{fig:stabilitylattice}b). 

In principle, any set of unit length wavevectors is permitted, though if $U$ is
required to be real, the negative of each vector must also be included. The
usual choice for periodic patterns is $Q=2$, 4 or~6 equally spaced wavevectors,
for stripes, squares or hexagons, though $Q=8$, 10 and~12 have also been used
in previous studies and have been observed in experiments. Synthetic examples
of patterns and quasipatterns with $A_j=\mbox{constant}$ for $Q=4,\dots,14$ are
shown in figure~\ref{fig:qpexamples}(d--i).

At each higher order in~$\epsilon$, the equation to solve takes the form
 \begin{equation}\label{eq:ordern}
 \calLz(U_n)=- \calNn(U_1,\dots,U_{n-1};\mu_1)
                     + \frac{\partial U_{n-1}}{\partial t},
 \end{equation}
where the term $\calNn$ is given by the order~$\epsilon^n$ part of
Taylor expansion of $\calFmu(U)$ in powers of~$\epsilon$, so it
contains nonlinear terms and the parameter~$\mu_1$ from~$\calLmu(U)$.
In principle, the equations can be solved order by order, with each $U_n$
determined by $U_1,\dots,U_{n-1}$. At each order, the nonlinear terms
$\calN_2$, $\calN_3$, etc., involve quadratic, cubic, etc.,
combinations of the original Fourier modes, which implies that these terms will
involve powers of (at most)~$n$ in the original amplitudes, and that the
Fourier spectrum of~$\calNn$ contains wavevectors made up (at most) of
all combinations of up to $n$ of the original wavevectors in the set~$K$. If we
let
 \begin{equation}
 \bfkm=\sum_{j=1}^Q m_j \bfk_j,
 \qquad\mbox{where}\qquad
 \bfm\in\Zset^Q,
 \end{equation}
then the wavevectors in the nonlinear terms at order~$n$ are all those with
$\bfkm$ satisfying $|\bfm|=\sum_j|m_j|\leq n$.
 
In order to solve the equation~(\ref{eq:ordern}) at order~$n$, the modes
present in $\calNn$ are divided into two classes. First, if a mode has
wavevector on the unit circle, then, using the orthogonality in~$\Rset^2$ of
Fourier modes with different wavevectors (or, more properly, using solutions of
the adjoint equation and integrating over~$\Rset^2$), the coefficient of this
mode on the RHS of~(\ref{eq:ordern}) must be zero (this condition is known as a
solvability condition). The reason for this is that $\calLz(\eikdotx)$
is zero if $|\bfk|=1$, and so such modes are not present in the LHS
of~(\ref{eq:ordern}). So, for example, the evolution of the amplitudes $A_j(t)$
is determined at second order and takes the general form:
 \begin{equation}\label{eq:evolutionleading}
 {\dot A_j} = f_j(A_1,\dots,A_Q;\mu_1),
 \end{equation}
where the dot stands for evolution on the slow time scale. These evolution
equations will contain only terms linear and quadratic in the $A$'s. In
examples with additional symmetry (or with $Q\neq6$ and $Q\neq12$ equally spaced
modes), this equation is vacuous, with $\mu_1=0$ and $U_2=0$, and the evolution
of the amplitudes $A_j(t)$ is determined at third order -- see
section~\ref{sec:SwiftHohenberg}.

Once the modes on the unit circle have been removed by satisfying a solvability
condition, all remaining modes in in $\calNn$ (making up the second
class of modes) have wavevectors off the unit circle. For these modes,
$\calLz(\eikdotx)$ is nonzero, so the singular linear operator operator
$\calLz$ can be inverted to give~$U_n$:
 \begin{equation}\label{eq:generalstep}
 U_n = \calLzinv\left(
 - \calNn(U_1,\dots,U_{n-1};\mu_1) + \frac{\partial U_{n-1}}{\partial t},
 \right).
 \end{equation}
Inverting the operator~$\calLz$ generates arbitrary linear combinations
of modes in its kernel that are used to satisfy solvability conditions at
higher order. Like the nonlinear term~$\calNn$, each $U_n$ will include
modes with wavevectors made up of (at most) of all combinations~$\bfkm$ of up
to $n$ of the original wavevectors, with~$|\bfm|\leq n$.

As well as the method outlined above, there are two other approaches to these
computations, both of which avoid adding modes on the unit circle at each order
when inverting~$\calLz$. First, the original paper~\cite{refM25} used an
expansion for the parameter~$\mu$: $\mu=\epsilon\mu_1+\epsilon^2\mu_2+\dots$,
and chose values of $\mu_1$, $\mu_2$, \hbox{etc.}, to satisfy the solvability
conditions at each order. As a result, for a given value of~$\mu$, a polynomial
for~$\epsilon$ must be solved before the amplitude can be computed. In the
second alternative, the original $\mu=\epsilon\mu_1$ is used, leaving modes in
the RHS of~(\ref{eq:ordern}) on the unit circle, which cannot be removed. The
modes involved will be exactly the $Q$~modes that were taken in the original
ansatz for~$U_1$, and the contributions that appear in~(\ref{eq:ordern}) can be
redesignated as order~$\epsilon^{n-1}$ corrections to the leading order
solvability condition~(\ref{eq:evolutionleading}). We have checked for some
specific examples that the three approaches give the same results, but prefer
the approach described in detail for the problem at hand.

In many cases, the leading order solvability
condition~(\ref{eq:evolutionleading}) is sufficient, but in other problems,
this equation is degenerate, and the calculation is carried to some higher
order~$N$ in powers of~$\epsilon$. An implicit {\em assumption} is that the
power series expansion~(\ref{eq:powerseriesU}) for~$U$ converges for some
nonzero~$\epsilon$ as this process is repeated (and~$N\rightarrow\infty$). At
each order~$n$ in the perturbation calculation, the operator~$\calLz$ must be
inverted for each mode~$\eikmdotx$, where $|\bfm|\leq n$ and $\bfkm$ could be
close to the unit circle. In typical pattern forming problems, the growth rate
of a mode~$\eikdotx$ has a quadratic maximum at $k=1$ (see
figure~\ref{fig:stabilitylattice}a), so
 \begin{equation}\label{eq:invertL0}
 \calLzinv\left(\eikdotx\right) 
  \approx -\frac{1}{(1-|\bfk|^2)^2}\eikdotx,
 \end{equation}
for $|\bfk|$ close to~1, with equality in the case of the Swift--Hohenberg
example. For periodic patterns (with $Q=2$, $4$ or $6$ modes), integer
combinations of the initial wavevectors form a lattice, so the
wavevectors~$\bfkm$ cannot come arbitrarily close to the unit circle (apart
from the modes on the unit circle, which are dealt with by applying solvability
conditions). In this case, convergence will not be a problem for small
enough~$\epsilon$. However, for quasipatterns (with $Q=8$, $10$, $12$ or more
modes), there is no lattice and combinations of modes can come arbitrarily
close to the unit circle, leading to {\em small divisors} in the denominator
when~$\calLz$ is inverted. The issue of convergence has never been properly
examined for two-dimensional quasipatterns, and most authors assume that the
leading order solvability condition~(\ref{eq:evolutionleading}) yields useful
and reliable information about the amplitude and stability of the quasipattern.
We will see below that it is far from obvious that this is the case.

This so-called small divisor problem is well known in other situations that
feature quasiperiodicity, and, in particular, is known to arise with these
quasipatterns. What is not known is how rapidly $\calLzinv$ grows as the order
of truncation~$N$ increases. The question here is how close $\bfkm$ can get to
the unit circle as $|\bfm|=N$ becomes large, and does the power series
expansion for~$U$ have a nonzero radius of convergence in spite of the bad
behaviour of~$\calLzinv$. We turn to these two questions in the next sections.

\section{Combinations of modes}
\label{sec:Modes}

In this section, we take integer combinations of up to $N$ of the $Q$~original
vectors on the unit circle, and compute how close these combinations can get to
the unit circle as $N$~becomes large. We are able to prove that the closest
$|\bfkm|$ can get to~1, with $|\bfm|=N$, is bounded above and below by a
constant times~$N^{-2}$ in the cases $Q=8$, 10 and~12. We also have numerical
evidence that the closest distance is bounded above and below by a constant
times~$N^{-4}$ in the case~$Q=14$ (at least for~$N\leq1000$), going to zero
much more rapidly than in the other three cases.

 \begin{figure}
 \begin{center}
 \mbox{\makebox[0.25\hsize][c]{(a)}
       \hspace{0.05\hsize}
       \makebox[0.25\hsize][c]{(b)}
       \hspace{0.05\hsize}
       \makebox[0.25\hsize][c]{(c)}}
 \mbox{\psfig{file=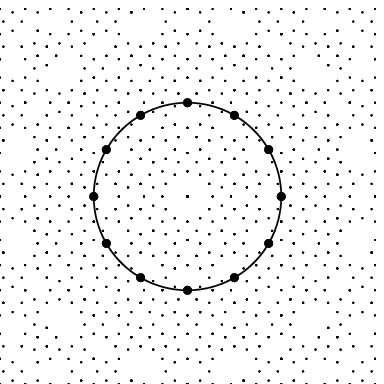,width=0.25\hsize}
       \hspace{0.05\hsize}
       \psfig{file=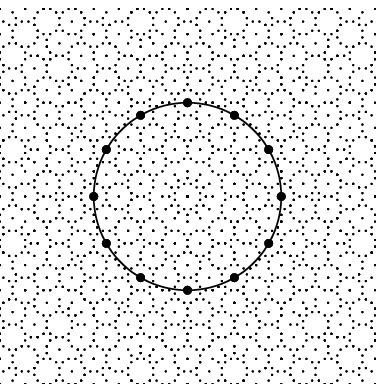,width=0.25\hsize}
       \hspace{0.05\hsize}
       \psfig{file=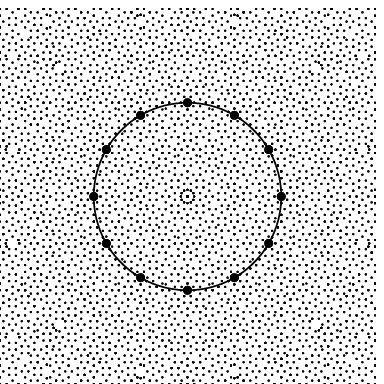,width=0.25\hsize}}
 \mbox{\makebox[0.25\hsize][c]{(d)}
       \hspace{0.05\hsize}
       \makebox[0.25\hsize][c]{(e)}
       \hspace{0.05\hsize}
       \makebox[0.25\hsize][c]{(f)}}
 \mbox{\psfig{file=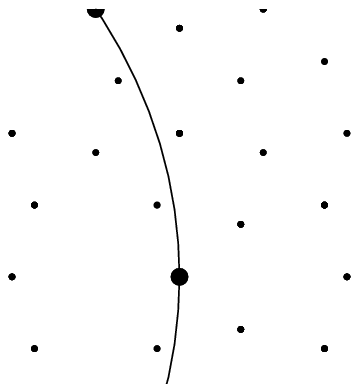,width=0.25\hsize}
       \hspace{0.05\hsize}
       \psfig{file=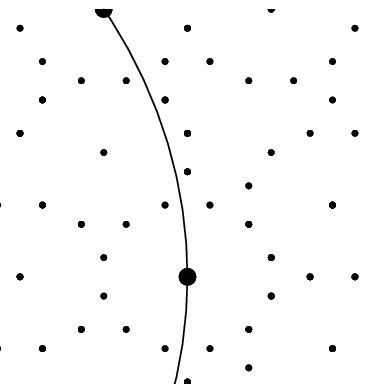,width=0.25\hsize}
       \hspace{0.05\hsize}
       \psfig{file=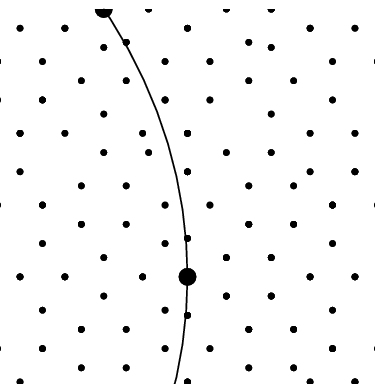,width=0.25\hsize}}
 \mbox{\makebox[0.25\hsize][c]{(g)}
       \hspace{0.05\hsize}
       \makebox[0.25\hsize][c]{(h)}
       \hspace{0.05\hsize}
       \makebox[0.25\hsize][c]{(i)}}
 \mbox{\psfig{file=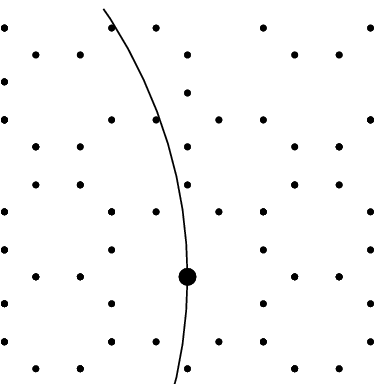,width=0.25\hsize}
       \hspace{0.05\hsize}
       \psfig{file=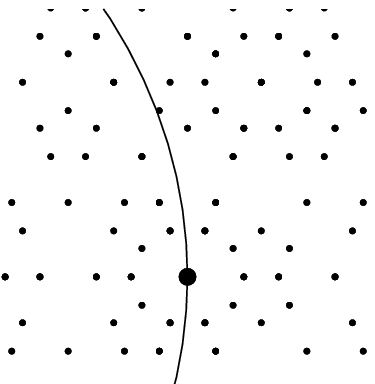,width=0.25\hsize}
       \hspace{0.05\hsize}
       \psfig{file=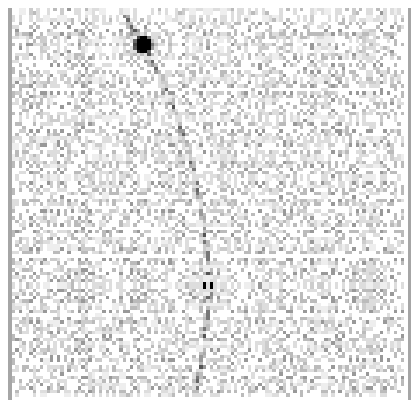,width=0.25\hsize}}
 \end{center}
 \caption{Positions of combinations of up to~$N$ wavevectors original vectors
on the unit circle, with (a)~$Q=12$, $N=7$, (b)~$Q=12$, $N=11$, (c)~$Q=12$,
$N=15$; (d,e,f) on second row: details of first row. The circle indicates the
unit circle,~$k=1$, the large dots are the original $Q$~wavevectors, and the
small dots are integer combinations of these. (g), (h) and (i)~show $N=15$ and
$Q=8$, 10 and 14. Note how the density of points increases with~$N$ and
with~$Q$, and the proximity of points to the unit circle decreases with~$N$.
Note also how the density of points is markedly higher with~$Q=14$, for the
same value of~$N$.}
 \label{fig:wavevectors}
 \end{figure}

 \begin{figure}
 \begin{center}
 \mbox{\psfig{file=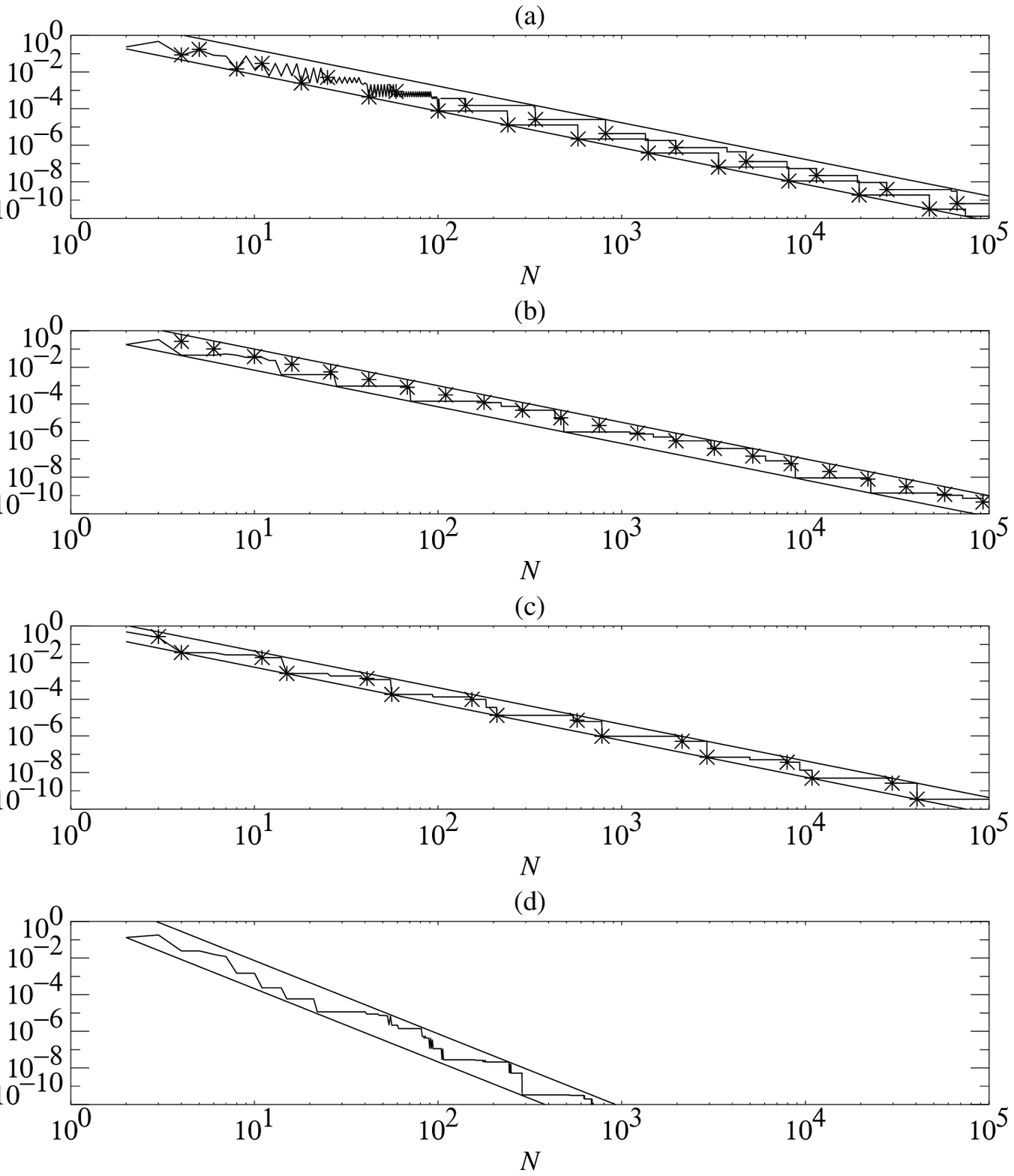,width=\hsize}}
 \end{center}
 \caption{Smallest nonzero distances from the unit circle
 $\left||\bfkm|-1\right|$
 as a function of the total number of modes $|\bfm|=N$, for (a)~$Q=8$,
 (b)~$Q=10$, (c)~$Q=12$ and (d)~$Q=14$.
 Stars in (a--c) mark distances calculated 
 from equations (\ref{eq:vectors08}--\ref{eq:vectors12}), 
 and straight lines indicate the scaling~$N^{-2}$.
 The two staircase-shaped lines in~(a) indicate minimum distances for even
 and odd values of~$N$. 
 The straight lines in~(d) indicate~$N^{-4}$.}
 \label{fig:closest}
 \end{figure}

We begin with figure~\ref{fig:wavevectors}(a--f), illustrating the locations of
combinations of up to $N=7$, 11 and 15 wavevectors in the case $Q=12$. Note how
the density of points increases with~$N$, and how the minimum distance between
points and the unit circle goes down with~$N$.
Figure~\ref{fig:wavevectors}(g--i) compares with the cases $Q=8$ and~10, which
show similar behaviour, and with $Q=14$, which has a much higher density of
possible wavevectors for the same value of~$N$.

Figure~\ref{fig:closest} shows detailed numerical results for the smallest
nonzero distance $\left||\bfkm|-1\right|$ from the unit circle as a function
of the total number of modes $|\bfm|=N$ for $Q=8$, 10, 12 and~14 original
modes. The calculations are made possible by a rapid method of searching for
the closest approach, presented in appendix~\ref{app:RapidMethod}: the method
is order~$N^2$ for $Q=8$, 10, 12 and order~$N^4$ for $Q=14$. The solid lines in
figure~\ref{fig:closest} confirm numerically that the scaling for the distance
to the unit circle is order~$N^{-2}$ for $Q=8$, 10 and~12, and order~$N^{-4}$
for $Q=14$. The remainder of this section is devoted to proving the correctness
of the $N^{-2}$~scalings, and in particular, to showing how for certain values
of~$N$, wavevectors close to the unit circle can be found explicitly in the
cases $Q=8$, 10 and~12, using continued fraction expansions.

We label the vectors $\bfk_1$, $\bfk_2$, \dots, $\bfk_Q$ anticlockwise around
the circle starting with $\bfk_1=(1,0)$, with $\bfk_{j+Q/2}=-\bfk_j$. We are
interested in the scaling behaviour of how close $\bfkm=\sum_{j=1}^Q m_j
\bfk_j$ can get to the unit circle as $|\bfm|=\sum_j|m_j|=N$ becomes large, so
we seek the vector of integers~$\bfm$ with $|\bfm|=N$ that yields the
vector~$\bfkm$ that is closest to the unit circle for this value of~$N$. Once
we have found a particular vector that is close to the unit circle, we are also
interested in finding the smallest~$N$ that can achieve this distance.
Including equal and opposite vectors $\bfk_j$ and $\bfk_{j+Q/2}$ will only
increase~$N$ without coming any closer to the unit circle, so we take only
$m_1,\dots,m_{Q/2}$ but allow these to be negative.

With this restriction, the squared length of a vector $\bfkm$ is, for each 
value of~$Q$:
 \begin{eqnarray*}
 Q=2:&\quad |\bfkm|^2 =& m_1^2\\
 Q=4:&\quad |\bfkm|^2 =& m_1^2 + m_2^2\\
 Q=6:&\quad |\bfkm|^2 =& m_1^2 + m_2^2 + m_3^2 + m_1m_2 + m_2m_3 - m_3m_1\\
 Q=8:&\quad |\bfkm|^2 =& m_1^2 + m_2^2 + m_3^2 + m_4^2 \\
     &                 &{}+ \sqrt{2}(m_1m_2 + m_2m_3 + m_3m_4 - m_4m_1)\\
Q=10:&\quad |\bfkm|^2 =& m_1^2 + m_2^2 + m_3^2 + m_4^2 + m_5^2 \\
     &                 &{}+ \omega(m_1m_2 + m_2m_3 + m_3m_4 + m_4m_5 - m_5m_1)\\
     &                 &{}+ (\omega-1)(m_1m_3 + m_2m_4 + m_3m_5 - m_4m_1 - m_5m_2)\\
Q=12:&\quad |\bfkm|^2 =& m_1^2 + m_2^2 + m_3^2 + m_4^2 + m_5^2 + m_6^2\\
     &&{}      + m_1m_3 + m_2m_4 + m_3m_5 + m_4m_6 - m_5m_1 - m_6m_2\\ 
     &&{}      + \sqrt{3}(m_1m_2 + m_2m_3 + m_3m_4 + m_4m_5 + m_5m_6 - m_6m_1)\\
Q=14:&\quad |\bfkm|^2 =& m_1^2 + m_2^2 + m_3^2 + m_4^2 + m_5^2 + m_6^2+ m_7^2\\
 &&{}+ \omega_1(m_1m_2 + m_2m_3 + m_3m_4 + m_4m_5 + m_5m_6 + m_6m_7 - m_7m_1)\\
 &&{}+ \omega_2(m_1m_3 + m_2m_4 + m_3m_5 + m_4m_6 + m_5m_7 - m_6m_1 - m_7m_2)\\ 
 &&{}+ \omega_3(m_1m_4 + m_2m_5 + m_3m_6 + m_4m_7 - m_5m_1 - m_6m_2 - m_7m_3)
 \end{eqnarray*}
where $\omega$ is the golden ratio: $\omega=(1+\sqrt{5})/2=2\cos(\pi/5)$, with
$\omega^2=\omega+1$, and $\omega_j=2\cos(j\pi/7)$, with $\omega_2=\omega_1^2-2$
and $\omega_3=1-\omega_1+\omega_2$. The irrational~$\omega_1$ is the root of a
cubic equation: $\omega_1^3-\omega_1^2-2\omega_1+1=0$.

We observe that for $Q=2$, 4 and~6, $|\bfkm|^2$ is an integer, so, as expected,
points on square and hexagonal lattices cannot come arbitrarily close to the
unit circle without actually hitting it.

For $Q=8$, 10 and~12, $|\bfkm|^2$ is of the form:
 \begin{equation}
 |\bfkm|^2 = 1 + p - rq,
 \end{equation}
where $r=\sqrt{2}$, $\omega$ or~$\sqrt{3}$ is an irrational root of a quadratic
equation with integer coefficients, and $p$ and $q$ are integers. If $p - rq$
is close to zero (that is, if $r$ is well approximated by $\frac{p}{q}$), then
$|\bfkm|^2$ can come close to~1. The particular rational approximations
involved are:
 \begin{eqnarray}\label{eq:partfracq8}
  Q=8:&\quad \sqrt{2} \approx&\frac{p}{q}=\frac{m_1^2 + m_2^2 + m_3^2 + m_4^2 - 1}
                                   {m_4m_1 - m_3m_4 - m_2m_3 - m_1m_2},\\
 \label{eq:partfracq10}
 Q=10:&\quad \omega\approx & \frac{p}{q}=\frac{m_1^2 + \dots + m_5m_2 - 1}
                                  {m_5m_2 - \dots - m_1m_2},\\
 \label{eq:partfracq12}
 Q=12:&\quad \sqrt{3} \approx&\frac{p}{q}=\frac{m_1^2 + \dots - m_6m_2 - 1}
                                   {m_6m_1 - \dots - m_1m_2}.
 \end{eqnarray}
In the expressions above, we choose $p$ and $q$, which depend on the
integers~$\bfm$, to be positive and to have no common factors. 

For $Q=14$ (and higher), $|\bfkm|^2$ involves the sum of an integer plus
integers times at least two different irrationals that will not, in general, be
roots of quadratic equations with integer coefficients. Apart from the
numerical evidence in figure~\ref{fig:closest}, we do not pursue the cases
$Q\geq14$ further here.

\begin{table}                     
\begin{center}
\small                       
\begin{tabular}{r|rrrrrrrrrrr}
\hline                           
          & $l=0$ & $1$ & $2$ & $3$ & $4$ & $5$ & $6$ & $7$ & $8$ & $9$ & $10$ \\ 
\hline              
$r=\sqrt{2}$ & $\frac{p_l}{q_l}=\frac{1}{1}$ & $\frac{3}{2}$ & $\frac{7}{5}$ & $\frac{17}{12}$ & $\frac{41}{29}$ & $\frac{99}{70}$ & $\frac{239}{169}$ & $\frac{577}{408}$ & $\frac{1393}{985}$ & $\frac{3363}{2378}$ & $\frac{8119}{5741}$ \\
\hline
$r=\omega$   & $\frac{p_l}{q_l}=\frac{1}{1}$ & $\frac{2}{1}$ & $\frac{3}{2}$ & $\frac{5}{3}$ & $\frac{8}{5}$ & $\frac{13}{8}$ & $\frac{21}{13}$ & $\frac{34}{21}$ & $\frac{55}{34}$ & $\frac{89}{55}$ & $\frac{144}{89}$ \\
\hline
$r=\sqrt{3}$ & $\frac{p_l}{q_l}=\frac{1}{1}$ & $\frac{2}{1}$ & $\frac{5}{3}$ & $\frac{7}{4}$ & $\frac{19}{11}$ & $\frac{26}{15}$ & $\frac{71}{41}$ & $\frac{97}{56}$ & $\frac{265}{153}$ & $\frac{362}{209}$ & $\frac{989}{571}$ \\
\hline
\end{tabular}
\vspace{2mm}
 \caption{Continued fraction approximations to $r=\sqrt{2}$, $\omega$ 
 and $\sqrt{3}$, as a function of the order~$l$ of the truncation.}
 \label{tab:ContinuedFractions}
\end{center}
\end{table}

It is clear that the theory of continued fraction approximations of irrationals
will be useful here. The continued fraction expressions for the irrationals
$r=\sqrt{2}$, $\omega$ and $\sqrt{3}$ are (respectively):
 \begin{equation}\label{eq:ContinuedFractionExamples}
 1+{1\over\displaystyle 2+
                {\strut 1\over\displaystyle 2+\cdots}},
 \quad
 1+{1\over\displaystyle 1+
                {\strut 1\over\displaystyle 1+\cdots}},
 \quad
 1+{1\over\displaystyle 1+
                {\strut 1\over\displaystyle 2+
                  {\strut 1\over\displaystyle 1+
                    {\strut 1\over\displaystyle 2+\cdots}}}}.
 \end{equation}
If these fractions are truncated after $l$ terms, the successive fractions
$p_l/q_l$ that approximate~$r$ are given in table~\ref{tab:ContinuedFractions}.
We recall from the theory of continued fractions~\cite{refH60} that 
 \begin{equation}\label{eq:CFone}
 \frac{K_1}{q_l^2} < \left| \frac{p_l}{q_l} - r\right| < \frac{K_2}{q_l^2}
 \end{equation} 
for $r=\sqrt{2}$, $\omega$, $\sqrt{3}$, and $K_1$, $K_2$ are constants. The
values of $K_1$ and~$K_2$, which are order unity, are related to the largest
integers appearing in the expansions~(\ref{eq:ContinuedFractionExamples}),
which are 1 or 2 in these cases. These inequalities mean that the truncated
continued fraction expansions $\frac{p_l}{q_l}$ approximate~$r$ well, but not
too well, as $l$~becomes large. It will also be useful to note that if $l>1$
and if $q$~is an integer with $0<q<q_l$, then~\cite{refH60}
 \begin{equation}\label{eq:CFtwo}
 \left|\frac{p_l}{q_l}-r\right| < \left|\frac{p}{q}-r\right|.
 \end{equation} 
This means that if $\frac{p_l}{q_l}$ is the truncation of the continued
fraction approximation of an irrational~$r$, no other fraction with a smaller
denominator comes closer to~$r$.

If we exclude those vectors~$\bfkm$ that fall exactly on the unit circle, which
would have~$p=q=0$, the relations in~(\ref{eq:CFone}) and (\ref{eq:CFtwo}) can
be used to show that $|\bfkm|^2$ can approach~1 no faster than order~$N^{-2}$
for $Q=8$, 10 and~12. The reason is that the denominators
in~(\ref{eq:partfracq8}--\ref{eq:partfracq12}) are of the form of a sum of
products of the integers~$m_1,\dots,m_{Q/2}$. Since $\sum_j|m_j|=N$, each
quadratic term in the denominator can be no larger than~$N^2$ in magnitude;
there are no more than~$Q$ of these terms, so the denominator as a whole
satisfies~$q\leq QN^2$. Then, using first~(\ref{eq:CFtwo}) and
then~(\ref{eq:CFone}), we have
 \begin{equation}
 \left||\bfkm|^2-1\right| = \left|p-rq\right| \geq \left|p_l-rq_l\right|
                                              > \frac{K_1}{q_l},
 \end{equation} 
where $q_l$ is the smallest of the $q_l$'s above~$q$: $q_{l-1}<q\leq q_l$
(unless $q_{l-1}=q_l=1$). Now for $r=\sqrt{2}$, $\omega$ and~$\sqrt{3}$, the
$q_l$'s are no further than a factor of~3 apart~\cite{refH60}, so $q_l\leq
3q_{l-1}<3q\leq3QN^2$, and we have (assuming $|\bfkm|\neq1$):
 \begin{equation}
 \left||\bfkm|^2-1\right| > \frac{K}{N^2},
 \end{equation} 
where $|\bfm|=N$ and $K$~is a constant.

We now show that the order~$N^{-2}$ rate of approach is indeed achieved for 
$Q=8$, 10 and~12.
Observe that, 
for $Q=8$:
 \begin{equation}\label{eq:vectors08}
 \bfkm = \bfk_1 + p_l \bfk_3 + q_l\bfk_6 + q_l\bfk_8 = (1, p_l-\sqrt{2}q_l),
 \end{equation}
with $|\bfm|=N=p_l + 2q_l + 1$ (even)
and $|\bfkm|^2-1=p_l^2 + 2q_l^2 - 2p_lq_l\sqrt{2}$
(a related vector with $N=2p_l + 2q_l + 1$ (odd) can also be found);
for $Q=10$:
 \begin{eqnarray}\label{eq:vectors10}
 \bfkm &=& (p_l+1) \bfk_2 + (p_l-1) \bfk_5 + (q_l+1) \bfk_8 + (q_l-1) \bfk_9\\
 \nonumber
     {}&=&(1,(p_l-\omega q_l)\sqrt{3-\omega}),
 \end{eqnarray}
with $|\bfm|=N=2p_l + 2q_l$
and $|\bfkm|^2-1=3p_l^2 + 2q_l^2 +2p_lq_l- (p_l^2-q_l^2+4p_lq_l)\omega$, using 
the fact that $\omega^2=\omega+1$; and
for $Q=12$:
 \begin{equation}\label{eq:vectors12}
 \bfkm = p_l \bfk_4 + (q_l-1)\bfk_9 + (q_l+1)\bfk_{11} = (1,p_l-\sqrt{3}q_l),
 \end{equation}
with $|\bfm|=N=p_l + 2q_l$ and $|\bfkm|^2-1=p_l^2 + 3q_l^2 - 2p_lq_l\sqrt{3}$.
These vectors were found after a prolonged examination of the distances plotted
in figure~\ref{fig:closest}.

Using~(\ref{eq:CFone}), we have, for these particular vectors,
 \begin{equation}
 \left||\bfkm|^2-1\right| = (p_l-rq_l)^2 \leq \frac{K_2^2}{q_l^2},
 \end{equation}
where $r$ stands for $\sqrt{2}$, $\omega$ or~$\sqrt{3}$,
with an extra factor of $3-\omega$ in the case $r=\omega$. Using relations like
$q_l\leq p_l\leq q_{l+1}\leq3q_l$~\cite{refH60}, and the relations
between~$N$, $q_l$ and~$p_l$ above, it is then possible to show in each case
that $N$~is less than a constant times~$q_l$, so
 \begin{equation}
 \left||\bfkm|^2-1\right| < \frac{K'}{N^2},
 \end{equation}
where~$K'$ is a constant. These particular choices of~$\bfkm$ are plotted on
the graphs in figure~\ref{fig:closest}(a--c) as stars. 

From the graphs it is clear that these particular vectors are not always the
closest ones that can be found for given values of~$N$ (particularly for
$Q=10$), but they suffice to prove the scaling results required here. If
$|\bfkm|^2-1$ as given above (that is,
$|\bfkm|^2-1=p_l^2+2q_l^2-2p_lq_l\sqrt{2}$ for $Q=8$, and so on) can be written
as $p_{l'}-rq_{l'}$ for some integer~$l'$, then we would expect $|\bfkm|^2$ to
be particularly close to~1, compared with other vectors of up to that order.
So, for example, an excursion into numerology suggests the following relations,
which we have proven for $Q=8$ by induction: for $Q=8$, $l'=2l+1$; and for
$Q=12$, $l'=2l+1$ if $l$~is odd. On the other hand, for $Q=10$,
$p_l^2-q_l^2+4p_lq_l$ does not appear to equal~$q_{l'}$ for values of~$l$ up
to~15, which is probably why the constructed vector does not achieve the
closest possible distance in this case.

The summary result of this section is that we have shown that, given an
integer~$N$, the vector~$\bfkm$ with $|\bfm|=N$ that comes closest to the unit
circle (without being on the unit circle) satisfies
 \begin{equation}\label{eq:summaryinequalities}
 \frac{K}{N^2} \leq \left||\bfkm|^2-1\right| \leq \frac{K'}{N^2},
 \end{equation}
for constants~$K$ and~$K'$, for $Q=8$, 10 and 12 equally spaced original
vectors. The numerical evidence, for $N\leq10^6$, suggests values for $Q=8$:
$K=0.72$ and $K'=16.95$; for $Q=10$: $K=0.69$ and $K'=9.94$; and for $Q=12$:
$K=0.56$ and $K'=4.34$.


It should be emphasised that several of the steps in the derivation of these
bounds (for example, equation~(\ref{eq:CFone})) rely on the fact that
$\sqrt{2}$, $\omega$ and~$\sqrt{3}$ are quadratic irrationals, that is, they
are roots of quadratic equations with integer coefficients. This implies that 
the integers in the continued fraction expansion of these
numbers~(\ref{eq:ContinuedFractionExamples}) form repeating sequences and so 
are bounded above. The case of $Q=14$ (and higher) is more difficult to analyse
because the irrational numbers~$\omega_1$ and~$\omega_2$ in the expression
for~$|\bfkm|^2$ in this case are not quadratic irrationals, and because there
are two irrationals.

\section{The question of convergence}
\label{sec:Convergence}

The results of the previous section, taken with~(\ref{eq:invertL0}), imply that when $\bfkm$ is close to the
unit circle, $\calLzinv(\eikmdotx)$ can be as large as a constant
times $N^4\eikmdotx$ for $Q=8$, 10 and~12, with~$N=|\bfm|$.

At each stage in the calculation, the linear operator~$\calLz$ is 
inverted as in~(\ref{eq:generalstep}):
 \begin{equation}
 U_n = \calLzinv\left(
 - \calNn(U_1,\dots,U_{n-1})\right),
 \end{equation}
where we simplify the discussion by dropping the parameter and the time
dependence, and we assume that the solvability condition has been satisfied so
that $\calLz$~can be inverted . The nonlinear terms $\calNn$
will contain modes $\eikmdotx$ with $|\bfm|$~up to and including~$n$, so, at
least at first glance, $U_n$ will be $n^4$ times larger than the product of
various combinations of $U_1$, \dots, $U_{n-1}$. In particular, the combination
$U_1U_{n-1}$ occurs in~$\calNn$. This suggests that $U_n$ is $n^4$ times
larger than~$U_{n-1}$, which is $(n-1)^4$ times larger than~$U_{n-2}$, and so
on -- so $U_n$ will be of order $(n!)^4$. The perturbation method should yield
the pattern~$U(x,y)$ as the limit:
 \begin{equation}\label{eq:Ulimit}
 U = \lim_{N\rightarrow\infty}\sum_{n=1}^N \epsilon^nU_n,
 \end{equation}
but if $U_n$ is as large as $(n!)^4$, the limit will not converge for 
nonzero~$\epsilon$. For the limit to converge, $U_n$ should be no larger than a 
constant raised to the power of~$n$.

In fact, the $(n!)^4$ estimate is unduly pessimistic, since there will be
cancellations. Moreover, if we focus on nonlinear interactions 
in~$\calNn$ that result in modes with wavevectors on the unit circle, 
the modes that
are close to the unit circle involved in these nonlinear interactions will
originate from~$U_{n/2}$, not from~$U_{n-1}$. The reason is that the terms
in~$\calNn$ involve products like $U_1U_{n-1}$, $U_2U_{n-2}$, \dots,
$U_{n/2}^2$. Now the term $U_1U_{n-1}$ cannot generate modes closer to the unit 
circle than the modes already in~$U_{n-1}$. Modes that are closer than any 
previous combination will come (if at all) from combinations like~$U_{n/2}^2$, 
and will involve the modes in $U_{n/2}$ that are close to the unit circle. This 
leads to an estimate of the form: $U_n$ will be $n^4$ times larger 
than~$U_{n/2}^2$, which will be $(n/2)^8$ times larger than~$U_{n/4}^4$, and so 
on. The resulting estimate for $U_n$ will not be as large as~$(n!)^4$, but it 
will still be larger than any constant raised to the power of~$n$.

Unfortunately, it is difficult to be more precise than this, but the arguments 
above certainly cast doubt on the convergence of the modified perturbation 
theory method as applied in this way to determine the amplitude of a 
quasipattern as a function of parameter. At best the series will be {\em 
asymptotic}, that is, not converging and yet still yielding useful information
when truncated at low order.

As a cautionary tale, we turn briefly to the asymptotic series representation
of the Stieltjes integral, as described in~\cite{refB126}:
 \begin{equation} 
 I=\int_0^\infty\frac{e^{-t}}{1+\epsilon t}\,dt
  =\int_0^\infty\left\{1-\epsilon t + \epsilon^2 t^2 - \dots\right\}e^{-t}\,dt
  =\sum_{n=0}^\infty(-1)^n n!\epsilon^n.
 \end{equation}
$I$~is a perfectly well defined integral, depending on a small positive
parameter~$\epsilon$. The fraction may be expanded formally as a power series
in~$\epsilon$, but this step is invalid, as the sum does not converge if
$t>1/\epsilon$, and the upper limit of the integral is $t=\mbox{infinity}$. The
formal power series can be integrated term by term, resulting in an infinite
sum for the value of~$I$ as a function of~$\epsilon$. This sum does not
converge for any nonzero~$\epsilon$, and yet, if the sum is truncated at a
fixed order~$N$, then there is a range of~$\epsilon$ for which the truncated
sum gives a reasonable approximation of~$I$~\cite{refB126}. Taking a larger~$N$
results in a smaller range of~$\epsilon$, closer to zero. On the other hand,
if~$\epsilon$ is fixed at a small number, then there is a a truncation~$N$ that
gives an estimate of~$I$ that is reasonably close to the correct value. The
truncation can be increased as~$\epsilon$ is taken to be smaller.

The danger is that the relation between the usable range of~$\epsilon$ and the 
level of truncation is not known in advance. The most severe truncation ($I=1$) 
is the safest, but loses useful information about the dependence of~$I$ 
on~$\epsilon$.

In the next section, we take the specific example of steady quasipattern
solutions of the cubic Swift--Hohenberg equation~(\ref{eq:SwiftHohenberg}). The
equivalent severe truncation would have the amplitude of the quasipattern
be~$\sqrt{\mu}$, which has been widely used by authors who are relying on the
small divisor problem alluded to above not rendering this truncation
meaningless.

\section{Example: the Swift--Hohenberg equation}
\label{sec:SwiftHohenberg}

In this section, we go through the details of deriving expressions for the
amplitudes in the specific example of steady solutions of the Swift--Hohenberg
equation~(\ref{eq:SwiftHohenberg}). This is one of the simplest pattern-forming
PDEs, and serves to illustrate the problem at hand.

For this presentation, we concentrate on the case $Q=2$, but we have carried
out the computations for $Q=2,\dots,12$ and up to
$33^{\mbox{$\scriptstyle{\mathrm{rd}}$}}$~order in~$\epsilon$. The symmetry
$U\rightarrow-U$ implies that all even terms $U_2$, $U_4$, etc., are absent,
that $\mu_1=0$, and that time should be scaled by~$\epsilon^{-2}$. To simplify
the presentation, we will only seek steady solutions, and so drop the time
derivative terms. By taking $\epsilon^2$ to be the bifurcation parameter, we
can set $\mu_2=1$ (taking $\mu_2$ to be positive since the bifurcation is
supercritical). The expansion is then
 \begin{equation}
 U = \epsilon U_1 + \epsilon^3 U_3 + \epsilon^5 U_5 + \ldots,\qquad
 \mu = \epsilon^2, 
 \end{equation}
with $\calLz(U) = - (1+\nabla^2)^2 U$. 

The leading order equation at order~$\epsilon$ is $\calLz(U_1)=0$, which is
solved by (for real~$U$ and with $Q=2$ modes, for clarity of exposition):
 \begin{equation}
 U_1(x,y)=A_1\eix + {\bar A_1}\emix,
 \end{equation}
where altering the phase of the complex amplitude~$A_1$ translates the 
pattern.

At third order in~$\epsilon$, the equation we have to solve is:
 \begin{equation} \label{eq:orderthree}
 \calLz(U_3) = - U_1 + U_1^3. 
 \end{equation}
Now $U_1^3=A_1^3\enix{3} + 3|A_1|^2A_1\eix + \mbox{c.c.}$ (where $\mbox{c.c.}$
stands for complex conjugate), and since $\calLz(U_3)$ cannot contain
$\eix$ and $\emix$, equation (\ref{eq:orderthree}) gives:
 \begin{equation}\label{eq:leadingbranchingeqn}
 0 =  A_1 - 3|A_1|^2A_1.
 \end{equation}
This equation would usually have a factor of~$\mu_2$ in front of the linear
term, but we have set this to~1. The nontrivial solution of this equation is
$A_1=1/\sqrt{3}$, where we may take $A_1$ to be real without loss of
generality. The operator $\calLz$ acts on a mode $\eikdotx$ as
$\calLz(\eikdotx)=-(1-k^2)^2\eikdotx$, so $\calLz(U_3)$ can be
inverted to give:
 \begin{equation}
 U_3=-\frac{1}{192\sqrt{3}}\left(\enix{3} + \emnix{3}\right)
     + A_3\eix + {\bar A_3}\emix,
 \end{equation}
where we have included an arbitrary combination of the original modes on the
unit circle (in the kernel of~$\calLz$), which will allow the
solvability condition to be satisfied at the next order.

At fifth order in~$\epsilon$, the equation we have to solve is:
 \begin{eqnarray} \label{eq:orderfive}
 \calLz(U_5) &=& - U_3 + 3U_1^2U_3\\
 \nonumber
   {} &=&
   -\frac{1}{192\sqrt{3}}\left(\eix+\enix{3}+\enix{5}\right)
   + A_3\enix{3}
   + (A_3 + {\bar A_3})\eix
   + \mbox{c.c.}
 \end{eqnarray}
The solvability condition can be satisfied by setting $A_3=1/384\sqrt{3}$, with
an arbitrary imaginary component (set to zero) that corresponds to a
small translation of the original pattern. With the~$\eix$ component removed,
$\calLz$ can be inverted to give~$U_5$.

\begingroup
\def\phm{\phantom{-}}          
\begin{table}                     
\begin{center}
\small                       
\begin{tabular}{l|llllll}
\hline                           
          & \hfil$Q=2$ & \hfil4 & \hfil6 & \hfil8 & \hfil10 & \hfil12\\
\hline 
$A_{ 3}$ &$\phm10^{ -2.11}$ &$\phm10^{-0.72}$ &$\phm10^{ 0.57}$ &$\phm10^{ 0.78}$ &$\phm10^{ 1.11}$ &$\phm10^{ 1.54}$\\
$A_{ 5}$ &   $-10^{ -3.40}$ &   $-10^{-0.95}$ &   $-10^{ 1.26}$ &   $-10^{ 2.37}$ &   $-10^{ 3.08}$ &   $-10^{ 3.86}$\\
$A_{ 7}$ &$\phm10^{ -5.61}$ &   $-10^{-2.67}$ &   $-10^{ 2.75}$ &$\phm10^{ 4.75}$ &$\phm10^{ 6.21}$ &$\phm10^{ 7.15}$\\
$A_{ 9}$ &$\phm10^{ -5.91}$ &$\phm10^{-1.17}$ &$\phm10^{ 3.70}$ &   $-10^{ 7.22}$ &   $-10^{ 9.58}$ &   $-10^{10.79}$\\
$A_{11}$ &   $-10^{ -7.16}$ &   $-10^{-1.36}$ &$\phm10^{ 5.30}$ &$\phm10^{ 9.97}$ &$\phm10^{12.95}$ &$\phm10^{14.40}$\\
$A_{13}$ &   $-10^{ -8.62}$ &   $-10^{-1.57}$ &   $-10^{ 6.30}$ &   $-10^{12.88}$ &   $-10^{16.31}$ &   $-10^{17.98}$\\
$A_{15}$ &$\phm10^{ -9.31}$ &$\phm10^{-1.22}$ &   $-10^{ 7.94}$ &$\phm10^{15.79}$ &$\phm10^{19.66}$ &$\phm10^{21.55}$\\
$A_{17}$ &   $-10^{-10.90}$ &   $-10^{-1.75}$ &$\phm10^{ 8.96}$ &   $-10^{18.71}$ &   $-10^{23.01}$ &   $-10^{25.14}$\\
$A_{19}$ &   $-10^{-11.65}$ &   $-10^{-1.31}$ &$\phm10^{10.64}$ &$\phm10^{21.61}$ &$\phm10^{26.34}$ &$\phm10^{28.74}$\\
$A_{21}$ &$\phm10^{-12.70}$ &$\phm10^{-1.24}$ &   $-10^{11.65}$ &   $-10^{24.52}$ &   $-10^{29.68}$ &   $-10^{32.38}$\\
$A_{23}$ &$\phm10^{-14.47}$ &$\phm10^{-1.96}$ &   $-10^{13.37}$ &$\phm10^{27.46}$ &$\phm10^{33.02}$ &$\phm10^{36.07}$\\
$A_{25}$ &   $-10^{-14.83}$ &   $-10^{-1.13}$ &$\phm10^{14.37}$ &   $-10^{30.48}$ &   $-10^{36.36}$ &   $-10^{39.80}$\\
$A_{27}$ &$\phm10^{-16.22}$ &$\phm10^{-1.34}$ &$\phm10^{16.12}$ &$\phm10^{33.64}$ &$\phm10^{39.71}$ &$\phm10^{43.57}$\\
$A_{29}$ &$\phm10^{-17.18}$ &$\phm10^{-1.29}$ &   $-10^{17.10}$ &   $-10^{36.93}$ &   $-10^{43.07}$ &   $-10^{47.35}$\\
$A_{31}$ &   $-10^{-18.10}$ &   $-10^{-1.01}$ &   $-10^{18.88}$ &$\phm10^{40.29}$ &$\phm10^{46.46}$ &$\phm10^{51.15}$\\
\hline
\end{tabular}
\vspace{2mm}
 \caption{Values of the coefficients~$A_n$ in~(\ref{eq:Uexpansion}). These
 data are also plotted in figure~\ref{fig:coefficients}}
 \label{tab:Coefficients}
\end{center}
\end{table}
\endgroup

 \begin{figure}
 \begin{center}
 \mbox{\psfig{file=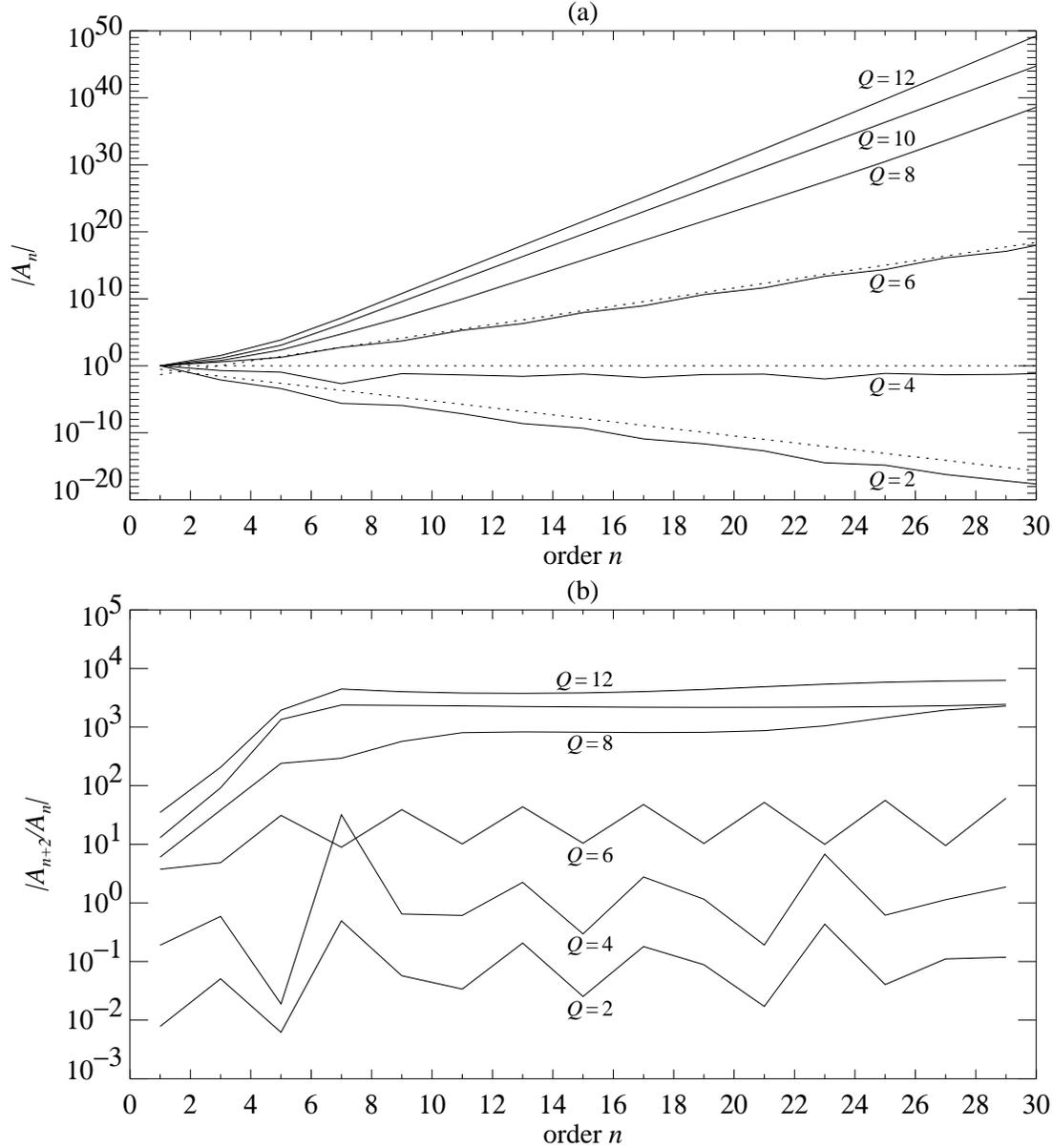,width=\hsize}}
 \end{center}
 \caption{(a)~Absolute values of the coefficients~$A_n$ in~(\ref{eq:Uexpansion})
 for $Q=2$ (lowest line), 4, 6, 8, 10 and 12 (top line) modes. These
 data are also given in table~\ref{tab:Coefficients}.
 Straight lines indicate that the coefficient of the modes on the unit
 circle will converge as the order of truncation~$N$ is increased, 
 and the inverse of the slope of the straight line gives the radius of
 convergence. The dotted lines are $0.3^n$ (lowest), $1^n$ and $4.8^n$
 (highest).
 (b)~Ratios $A_{n+2}/A_n$ against~$n$: this ratio continues to increase
 with~$n$ for $Q\geq8$, indicating that the sum
 in~(\ref{eq:AmplitudeTruncated}) may not converge for 
 nonzero~$\epsilon$.}
 \label{fig:coefficients}
 \end{figure}

This procedure is repeated up to some order~$N$, resulting in a power series
for the original pattern~$U$:
 \begin{eqnarray}
 \nonumber
 U &=& \left(
       \frac{\epsilon}{\sqrt{3}}
     + \frac{1}{128}   \left(\frac{\epsilon}{\sqrt{3}}\right)^3
     - \frac{13}{32768}\left(\frac{\epsilon}{\sqrt{3}}\right)^5 +\dots 
       \right) \left(\eix+\emix\right)\\
 \nonumber
 &&{}+\left(
      -\frac{1}{64}  \left(\frac{\epsilon}{\sqrt{3}}\right)^3
     + \frac{3}{8192}\left(\frac{\epsilon}{\sqrt{3}}\right)^5 +\dots
      \right)\left(\enix{3}+\enix{-3}\right)+\dots\\
 \nonumber
 &&{}+\left(
       \frac{1}{12288}\left(\frac{\epsilon}{\sqrt{3}}\right)^5 +\dots
      \right)\left(\enix{5}+\enix{-5}\right)+\dots\\
 \label{eq:Uexpansion}
   &=& \left(
       \sum_{n=1,3,\dots}^N
       A_n\left(\frac{\epsilon}{\sqrt{3}}\right)^n
       \right) \left(\eix+\emix\right) + \dots
 \end{eqnarray}
where $A_n$ is the coefficient of $(\epsilon/\sqrt{3})^n$ in the expression for
the amplitude of the modes on the unit circle. The factor~$\sqrt{3}$ is chosen
so that $A_1=1$ (so the $A_1$ and $A_3$ here are rescaled from the~$A_1$
and~$A_3$ above). The calculation can be carried out for other values of~$Q$;
for $Q=4,6,\dots,12$, the scaling for amplitudes is
$\sqrt{9},\sqrt{15},\dots,\sqrt{33}$, so $A_1=1$ in all cases. The values of
the first few coefficients~$A_n$ are given in table~\ref{tab:Coefficients} and
plotted in figure~\ref{fig:coefficients}. 

In these calculations, all modes generated by nonlinear interactions were kept
for $Q=2$, 4 and~6. For the other three cases, we kept only modes with
wavenumbers up to~$\sqrt{5}$, to keep the total number of modes within
manageable limits. Even so, with $Q=12$, there were more than 15000 modes
generated at the highest order -- without this truncation, there would have
been almost 2~million. We checked the lower orders against calculations keeping
all modes, and the differences in the mode amplitudes were about~1\%.
Restricting the number of modes in this way had no effect on how close
combinations of wavevectors could get to the unit circle.

For $Q=2$, 4 and~6, it appears that there will be no problem with the
convergence of the coefficient of the modes on the unit circle
in~(\ref{eq:Uexpansion}) since the coefficients~$A_n$ grow no faster than a
constant to the power of~$n$ (indicated by straight dotted lines in
figure~\ref{fig:coefficients}a). The bound on the ratios~$A_{n+2}/A_n$ also
indicates the range of values of~$\epsilon$ for which convergence is expected.

For $Q=8$, 10 and~12, the situation is less clear. The values of the
coefficients in table~\ref{tab:Coefficients} are certainly very large
($10^{50}$), but they are multiplied by the small parameter~$\epsilon$ raised
to high powers. At first glance, the values of the coefficients in
figure~\ref{fig:coefficients}(a) appear to be going up as straight lines (which
would indicate convergence), but the ratios~$A_{n+2}/A_n$ in
figure~\ref{fig:coefficients}(b) show that the slopes of these lines plateaus
for~$n$ in the range 12--18, and then start to increase gradually for
larger~$n$, suggesting that the coefficients~$A_n$ are growing faster than a
constant raised to the power of~$n$, and casting doubt on the convergence
of~(\ref{eq:Uexpansion}) as the level of truncation is increased.

The fact that the ratios~$A_{n+2}/A_n$ plateau and then start climbing in
figure~\ref{fig:coefficients}(b) can be related to the nonlinear interaction of
modes at each order in the perturbation theory. Taking $Q=8$ to be specific,
for $7\leq n\leq9$ and for $11\leq n\leq17$, the values of $|\bfkm|^2$
generated up to these orders, closest to the unit circle, are 0.85786 and
1.05887 respectively. Even though these modes close to the unit circle are
generated at some order~$n$, they do not influence the value of the coefficient
of modes on the unit circle until nonlinear interactions work their way back:
this will occur at order~$2n+1$. As a result, we expect the jump in~$k^2$ from
$n=9$ to~11 to influence~$A_{23}$ -- indeed, the ratios start to climb
at~$A_{23}$, and more sharply at~$A_{25}$. We would then expect these ratios to
plateau and then start climbing at order~39, plateau and climb again as
wavevectors generated at higher order come closer to the unit circle. The same
reasoning would predict the ratios to start climbing at $n=27$ for $Q=10$ and
at $n=15$ and 23 for $Q=12$, at least roughly consistent with the data in
figure~\ref{fig:coefficients}(b). It was only by going to such high order that
these issues became clear.

 \begin{figure}
 \begin{center}
 \mbox{\psfig{file=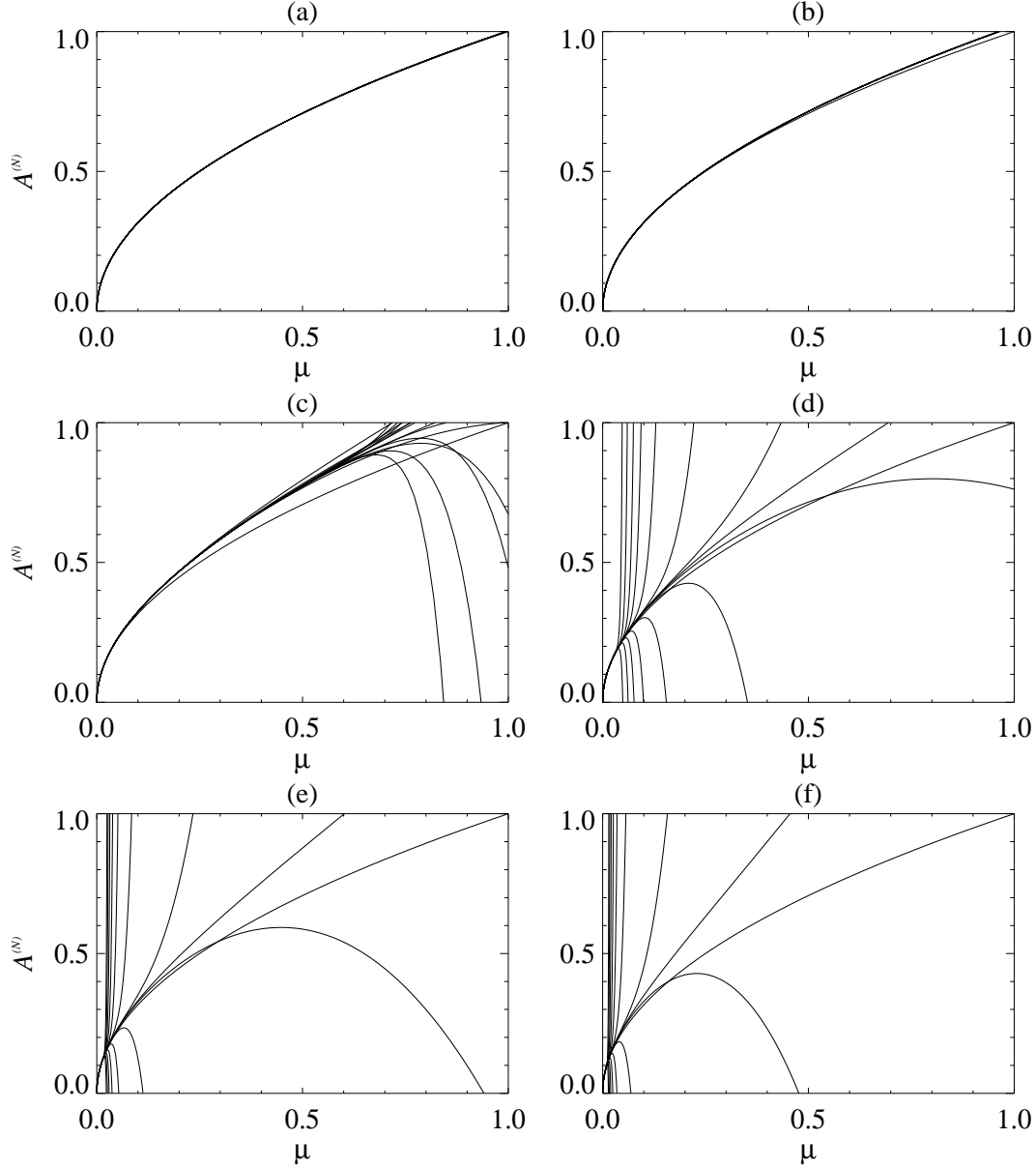,width=\hsize}}
 \end{center}
 \caption{Scaled amplitude~$A^{(N)}$ as a function of~$\mu$ 
 from~(\ref{eq:AmplitudeTruncated}), for 
 (a)~$Q=2$,
 (b)~$Q=4$,
 (c)~$Q=6$, 
 (d)~$Q=8$, 
 (e)~$Q=10$, 
 (f)~$Q=12$, 
 for different levels of truncation $N=1,\dots,31$.
 For $Q=2$ and $Q=4$, increasing the order of truncation
 has little effect for $\mu$~up to~1,
 while for $Q=6$, it appears that increasing the 
 order of truncation converges to a solution only for $\mu<0.65$ or so.
 However, for $Q\geq8$, increasing the order of truncation leads to graphs of
 $A^{(N)}$~as a function of~$\mu$ that appear to diverge for $\mu$~closer and
 closer to zero as $N$~becomes larger. Details of (c--f) are shown in 
 figure~\ref{fig:branchingdetail}.}
 \label{fig:branching}
 \end{figure}

 \begin{figure}
 \begin{center}
 \mbox{\psfig{file=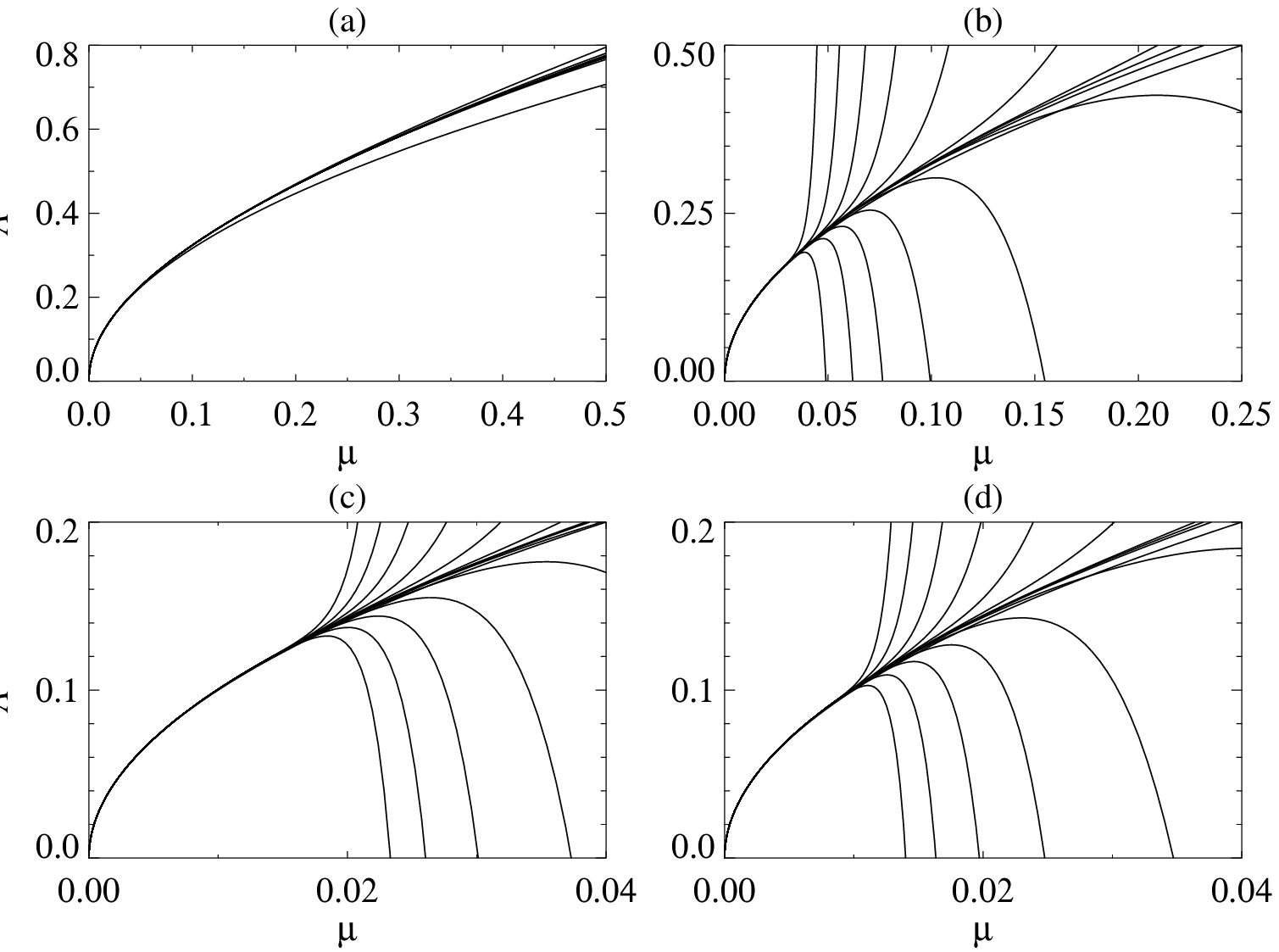,width=\hsize}}
 \end{center}
 \caption{Detail of figure~\ref{fig:branching}, for
 (a)~$Q=6$, 
 (b)~$Q=8$, 
 (c)~$Q=10$, 
 (d)~$Q=12$.
 Note how for $Q\geq8$, there is no sign that the graphs of
 $A^{(N)}$ against~$\mu$ are settling down as~$N$ increases.}
 \label{fig:branchingdetail}
 \end{figure}

Equation~(\ref{eq:Uexpansion}) can be used to find the amplitude~$A^{(N)}$ of
modes on the unit circle as a function of the original
parameter~$\mu=\epsilon^2$ when the expression is truncated to include powers
in~$\epsilon$ up to and including~$\epsilon^N$:
 \begin{equation}\label{eq:AmplitudeTruncated}
 A^{(N)} = \sqrt{\mu}\sum_{n=1,3,\dots}^N A_n\mu^{(n-1)/2},
 \end{equation}
where the amplitude has been rescaled (as indicated above) so that
$A^{(1)}=\sqrt{\mu}$ for all values of~$Q$. Graphs of $A^{(N)}$ as functions
of~$\mu$ for $N=1$ to~31 and for $Q=2$ to~12 are presented in
figures~\ref{fig:branching} and~\ref{fig:branchingdetail}. For $Q=2$, 4 and~6
(figure~\ref{fig:branching}a--c and figure~\ref{fig:branchingdetail}a), the
curves of~$A^{(N)}$ against~$\mu$ converge as~$N$ increases in a manner
consistent with the straight line increases of~$A_n$ in
figure~\ref{fig:coefficients}(a). In fact, an estimate of the radius of
convergence can be made from the slopes: the series will converge for
$\mu\leq32$ for $Q=2$, $\mu\leq9$ for $Q=4$, and $\mu\leq0.65$ for $Q=6$. These
limits are roughly half the values of $\mu$ at which modes generated in
nonlinear interactions become linearly unstable.

On the other hand, for $Q\geq8$ (figure~\ref{fig:branching}d--f and
figure~\ref{fig:branchingdetail}b--d), at each level of truncation~$N$, the
graph of~$A^{(N)}$ against~$\mu$ diverges at a value of~$\mu$ that is decreases
as~$N$ becomes larger, consistent with the steady increase in the
ratio~$A_{n+2}/A_n$ as $n$~increases. Indeed, modes generated in nonlinear
interactions become linearly unstable for~$\mu$ arbitrarily close to zero, for
large enough~$N$. However, it appears from these graphs that for
$\mu$~sufficiently small (say, less than~0.01 for~$Q=8$), there is a chance of
convergence. Without proper estimates of the rate of increase of~$A_n$
with~$n$, it is impossible to know for certain whether or not this method
converges, and if it does not, whether or not the truncated estimates converge
to the true solution in the limit of small~$\epsilon$.

\section{Discussion}
\label{sec:Discussion}

In summary, we have shown that modes generated by nonlinear interactions
between $Q=8$, 10 and 12 Fourier modes with wavevectors equally spaced around
the unit circle have wavevectors that can approach the unit circle no faster
than a constant times~$n^{-2}$, where $n$~is the number of modes involved. We 
have also shown by construction that there are combinations of modes that do
achieve this limit.

When carrying out modified perturbation theory in order to compute the
amplitude of a pattern as a function of a parameter, the usual approach is to
start with two assumptions: first, that when the parameter is small, the
desired pattern~$U$ can be written as a power series in that small parameter;
and second, the primary modes of interest have wavevectors equally distributed
around the unit circle. At each order~$n$ in the theory, nonlinear terms
generate modes involving up to~$n$ of the $Q$~modes. The modes that fall
exactly on the unit circle are dealt with by applying a solvability condition,
while equations for modes off the unit circle are satisfied by inverting the
linear operator~$\calLz$. In the cases $Q=2$, 4 and~6, the patterns are
spatially periodic and modes generated by nonlinear interactions do not
approach the unit circle. For $Q=8$, 10 and~12, the wavevector can come
within~$n^{-2}$ of the unit circle, and small divisors (order~$n^{-4}$) appear
when inverting~$\calLz$, leading to numerically large coefficients in
front of the Fourier modes. These coefficients grow sufficiently rapidly 
with~$n$ that convergence of the power series for the pattern~$U$ is called
into question.

We have explicitly carried out modified perturbation theory up to
$33^{\mbox{$\scriptstyle{\mathrm{rd}}$}}$~order for the cubic Swift--Hohenberg
equation. Of course, this kind of calculation cannot demonstrate convergence or
otherwise, but it does illustrate the issues that arise. The main conclusion of
the calculation is that even if modified perturbation theory does generate a
convergent series approximation to the quasipattern for small enough~$\mu$, the
series certainly diverges if the parameter~$\mu$ is bigger than about~$0.01$,
depending on exactly which value of~$Q$ is used. It is possible that the series
do converge for smaller~$\mu$, though we have argued that this is not the case.
Even if the series do diverge for all nonzero~$\mu$, a low-order truncation may
still give a useful approximation of the quasipattern, assuming that the
equations do have a quasipattern solution. It is on this basis that other
researchers have proceeded.

There are two related issues at stake. First, existence: do pattern forming
PDEs like the 2-dimensional Swift--Hohenberg equation have quasipattern
solutions that bifurcate from the trivial solution? Second, given the small
divisor problem, can asymptotic methods like modified perturbation theory yield
useful approximations to these solutions? We have not addressed the first issue
in this paper, but plan to turn to it in future. The limits we have derived on
the rate of approach of wavevectors to the unit circle will play a central role
in that calculation. As for the second issue, we have shown that modified
perturbation theory does not converge sufficiently rapidly (or slowly) to
provide an answer unequivocally one way or the other, and so this standard
method should not be regarded as a reliable way of computing properties of
quasipatterns.

What is needed is a method that converges more rapidly. Each order in the
standard theory gains a factor of~$\epsilon$ as well as large factors from any
small divisors that arise. There are other methods, developed for proofs of KAM
theory (see~\cite{refM125}) that converge more rapidly, and these may
be required for a rigorous treatment of quasipatterns as well. The difference
between the KAM situation and that of quasipatterns is that in the KAM case,
the solutions of interest are quasiperiodic in only one dimension (time), while
in the second, quasipatterns are quasiperiodic in two space directions.

By making arbitrarily small perturbations to the $Q$~wavevectors, it is
possible to make qualitative alterations to the nature of the problem in the
cases $Q=8$, 10 and~12. For instance, the patterns can be made periodic on
square or hexagonal lattices, with a lower limit to how close vectors can get
to the unit circle. For example, in the case $Q=12$, choosing the modes
$(1,0)$, $(\frac{2p_lq_l}{p_l^2+q_l^2},\frac{p_l^2-q_l^2}{p_l^2+q_l^2})$ and so
on. where $\frac{p_l}{q_l}$ is a continued fraction approximation to~$\sqrt{3}$
(see table~\ref{tab:ContinuedFractions}) yields 12~modes on the unit circle
that become nearly equally spaced as $l$~increases, and that generate a square
lattice by virtue of $(p_l^2-q_l^2,2p_lq_l,p_l^2+q_l^2)$ being Pythagorean
triplets -- see~\cite{refD54} for more details. Similarly, 12-dimensional
representations of the group~$D_6\sdp T^2$ can be chosen so that the modes are
nearly equally spaced and yet they generate a hexagonal lattice~\cite{refS105}.
Even in a square periodic domain, approximate $Q=12$ quasipatterns can be
generated~\cite{refS110}. The 8-dimensional representations of $D_4\sdp T^2$
can be used to approximate 8-fold quasipatterns in the same ways, though it is
not clear how a 10-fold quasipattern could be approximated by a periodic
pattern. The drawback with approximating quasipatterns by periodic patterns in
these ways is that the range of validity of the normal forms derived shrinks to
zero as the approximation improves.

It is interesting to note that 8, 10 and 12-fold quasipatterns have been
observed experimentally for several years now, but no 14-fold (or higher)
quasipattern has been reported (cf.~\cite{refN18}), with the possible exception 
of~\cite{refP54}. One might speculate that the reason for this is that the
convergence issues discussed above are likely to be more serious in the case
$Q=14$ since wavevectors approach the unit circle much more rapidly than in the
cases 8, 10 and~12 (see figure~\ref{fig:closest}).

 \begin{ack}
 We are grateful to many people who have helped shape these ideas, in one way
or another, over a period of several years: Peter Ashwin, Jon Dawes, Jay 
Fineberg, Rebecca Hoyle, Edgar Knobloch, Paul Matthews, Ian Melbourne, Michael
Proctor, Hermann Riecke and Mary Silber. The research of AMR is supported by
the Engineering and Physical Sciences Research Council.
 \end{ack}

\appendix

\section{Appendix: method for finding the closest modes}
\label{app:RapidMethod}

In this appendix, we present an order~$N^2$ algorithm for finding which
combinations of up to~$N$ vectors end up near the unit circle. The method is
suitable for $Q=8$, $10$ and $12$, and can be extended to an order~$N^4$ method 
for $Q=14$. We focus on $Q=12$ for definiteness, and let $\bfk_1=(1,0)$,
$\bfk_2=(\cos(2\pi/12),\sin(2\pi/12))$, etc.

For each value of~$N$, we want to find non-negative integers $m_j$ ($j=1$,
\dots, $12$) such that $\bfkm=\sum_j m_j\bfk_j$ is close to the unit circle,
with $|\bfm|=\sum_j|m_j|=N$, and $\bfm$ achieves this minimum distance to the
unit circle for this~$N$. In fact, we are interested in all combinations with
$|\bfm|\leq N$ satisfying this minimality condition. The requirement of
minimality and the symmetries of the problem lead to restrictions on the
integers $m_j$ that allow the order~$N^2$ algorithm

By rotating the vectors, we may choose $m_1>0$, without loss of generality.

The requirement for minimality amounts to considering only those $m_j$ where
there is no set of values $m'_j$ such that $\sum_j m_j \bfk_j = \sum_j m'_j
\bfk_j$ and $\sum|m'_j|<\sum|m_j|$. Using mod~12 arithmetic, this leads to:
 \begin{itemize}
 \item
 If $m_j>0$ then $m_{j+6}=0$, since otherwise let
   $m'_j=m_j-m_{j+6}$ and
   $m'_{j+6}=0$ (or the other way round if $m_{j+6}>m_j$), with 
   $m'_l = m_l$ for $l\neq j$, $j+6$. This
 gives a smaller set of vectors summing to the same point.
 \item
 If $m_j>0$ then $m_{j+4}=0$, 
  since $\bfk_j+\bfk_{j+4} = \bfk_{j+2}$, and we can let:
   $m'_{j+2} = m_{j+2} + \min(m_j, m_{j+4})$
   $m'_j = m_j - \min(m_j, m_{j+4})$
   $m'_{j+4} = m_{j+4} - \min(m_j, m_{j+4})$
   $m'_l = m_l$ for $l\neq j$, $j+2$, $j+4$.
 \item
 Similarly, if $m_j>0$ then $m_{j+8}=0$.
 \end{itemize}
Combined with $m_1>0$, these imply $m_5=m_7=m_9=0$. Also, only one of $m_3$ and
$m_{11}$ can be nonzero, by the same arguments. Using mirror symmetry, choose
$m_3>0$ and $m_{11}=0$.

Similar arguments applied to $m_2$, $m_4$, $m_6$, $m_8$, $m_{10}$ and $m_{12}$
imply that only two of these can be nonzero, and those two must be separated
by 2: $m_2$ and $m_4$, or $m_4$ and $m_6$, or $m_6$ and $m_8$, or $m_8$ and
$m_{10}$, or $m_{10}$ and $m_{12}$, or $m_{12}$ and $m_2$.
Consider these in turn, with $m_1>0$ and $m_3>0$.
 \begin{itemize}
 \item
 $m_2>0$ and $m_4>0$: if these two are nonzero, then all possible
     values of $\bfkm=\sum_j m_j \bfk_j$ are in the upper right 
     quadrant and cannot be close to the unit circle.
 \item
 $m_4>0$ and $m_6>0$: all possible values of $\bfkm$ lie in the upper
     half-plane, and cannot be close to the unit circle.
 \item
 $m_{10}>0$ and $m_{12}>0$, or $m_{12}>0$ and $m_2>0$: 
     all possible values of $\bfkm$ lie in the right
     half-plane, and cannot be close to the unit circle.
 \end{itemize}

The only remaining possibilities are either $m_6>0$ and $m_8>0$, or $m_8>0$ and
$m_{10}>0$. However, for every combination of $m_j$ in one configuration
$(m_1,m_3,m_6,m_8)$, there is an equivalent combination in the other
configuration $(m_1,m_3,m_8,m_{10})$ that has equal distance to the unit
circle: $m'_1=m_3$, $m'_3=m_1$, $m'_6=m_{10}$, $m'_8=m_8$, $m'_{10}=m_6$.
So we need only consider cases where $(m_1,m_3,m_8,m_{10})$ are nonzero.

Looping over all possible combinations of $(m_1,m_3,m_8,m_{10})$ with
$m_1+m_3+m_8+m_{10}\leq N$ gives an order $N^4$ algorithm, but this can be
improved as follows.

The vectors $\bfk_1=(1,0)$ and $\bfk_{10}=(0,-1)$.  If $\bfkm$ is to be close
to the unit circle, $m_3\bfk_3+m_8\bfk_8$ must lie in or near the upper left
quadrant of the wavevector plane (positive~$k_y$, negative~$k_x$), or at least
within the range $k_x<2$ and $k_y>-2$. Furthermore, for given $m_3$ and $m_8$,
the values of $m_1$ and $m_{10}$ for which~$\bfkm$ can lie close to the unit
circle are quite restricted: $m_1\bfk_1+m_{10}\bfk_{10}=(m_1,-m_{10})$ must be
near the vector $-m_3\bfk_3-m_8\bfk_8$.

So instead of looping over all possible combinations of $(m_1,m_3,m_8,m_{10})$,
it is only necessary to loop over $(m_3,m_8)$ and check values of $m_1$ close
to (within 2~of) the negative of the $x$~component of $m_3\bfk_3+m_8\bfk_8$,
and values of $m_{10}$ close to (within 2~of) the $y$~component of
$m_3\bfk_3+m_8\bfk_8$, which results in an order~$N^2$ algorithm.

The algorithm can be tidied up a little, and similar arguments can be applied
in the cases $Q=8$ and (with a little more difficulty) $Q=10$. When $Q\geq14$,
only order $N^4$ or slower algorithms are possible, based on the same ideas.
These methods were used to generate the data in figure~\ref{fig:closest}.


\begin{thebibliography}{99}



\def\JFM{J.~Fluid Mech.}
\def\GAFD{Geophys. Astrophys. Fluid Dynamics}

\def\ARFM{Annu. Rev. Fluid Mech.}
\def\PT{Phil. Trans. R.~Soc. Lond.~A}
\def\PRS{Proc. R.~Soc. Lond.~{\rm A}}
\def\PNASUSA{Proc. Nat. Acad. Sci. U.S.A.}

\def\PRL{Phys. Rev. Lett.}
\def\PRA{Phys. Rev.~A}
\def\PRE{Phys. Rev.~E}
\def\PLA{Phys. Lett.~A}
\def\JMP{J.~Math. Phys.}
\def\JPA{J.~Phys.~A}
\def\PCPS{Proc. Camb. Phil. Soc.}
\def\CRASP{C.~R. Acad. Sc. Paris}
\def\CUP{Cambridge University Press}

\def\JSP{J.~Stat. Phys.}

\def\IntJBifnChaos{Int.~J. Bifurcation and Chaos}

\def\SIAMJMA{SIAM J.~Math. Anal.}
\def\SIAMJAM{SIAM J.~Appl. Math.}
\def\SIAMJNA{SIAM J.~Num. Anal.}

\def\author#1#2{#1 #2}%

\def\refarticle#1#2#3#4#5#6#7{%
\bibitem{#1} #2, #4, {\it #5} {\bf #6} (#3) {#7}.}

\def\refarticleshort#1#2#3#4#5{%
\bibitem{#1} #2, #4, {\it #5} (#3).}

\def\refbook#1#2#3#4#5#6{%
\bibitem{#1} #2, {\it #4} (#5, #6, #3).}

\def\refartbook#1#2#3#4#5#6#7#8#9{%
\bibitem{#1} #2, #4, in: {#6}, eds., {\it #5} (#7, #8, #3) #9.}


\refarticle{refC119}
{\author{M.C.}{Cross} \and \author{P.C.}{Hohenberg}}{1993}
            {Pattern formation outside of equilibrium}
            {Rev. Mod. Phys.}{65}{851--1112}

\refbook{refR70}
{\author{M.I.}{Rabinovich}, \author{A.B.}{Ezersky} \and
 \author{P.D.}{Weidman}}{2000}
    {The Dynamics of Patterns}
    {World Scientific}{Singapore}

\refarticle{refA65}
{\author{H.}{Arbell} \and \author{J.}{Fineberg}}{2002}
            {Pattern formation in two-frequency forced parametric waves}
            {\PRE}{65}{036224}

\refbook{refG59}
{\author{M.}{Golubitsky}, \author{I.}{Stewart} \and \author{D.G.}{Schaeffer}}
    {1988}
    {Singularities and Groups in Bifurcation Theory. Volume~II}
    {Springer}{New York}

\refbook{refC23}
{\author{J.}{Carr}}{1981}
    {Applications of Centre Manifold Theory}
    {Springer}{New York}

\refbook{refJ41}
{\author{C.}{Janot}}{1994}
    {Quasicrystals: a Primer, 2nd edition}
    {Clarendon Press}{Oxford}


\refarticle{refC118}
{\author{B.}{Christiansen}, \author{P.}{Alstr{\o}m} \and 
 \author{M.T.}{Levinsen}}{1992}
            {Ordered capillary-wave states: quasicrystals, hexagons and
             radial waves}
            {\PRL}{68}{2157--2160}

\refarticle{refE15}
{\author{W.S.}{Edwards} \and \author{S.}{Fauve}}{1993}
            {Parametrically excited quasicrystalline surface waves}
            {\PRE}{47}{R788--R791}

\refarticle{refE10}
{\author{W.S.}{Edwards} \and \author{S.}{Fauve}}{1994}
            {Patterns and quasi-patterns in the Faraday experiment}
            {\JFM}{278}{123--148}

\refarticle{refK67}
{\author{A.}{Kudrolli}, \author{B.}{Pier} \and \author{J.P.}{Gollub}}{1998}
            {Superlattice patterns in surface waves}
            {Physica}{123D}{99--111}

\refarticle{refB90}
{\author{D.}{Binks} \and \author{W.}{van de Water}}{1997}
            {Nonlinear pattern formation of Faraday waves}
            {\PRL}{78}{4043--4046}

\refarticle{refB91}
{\author{D.}{Binks}, \author{M.-T.}{Westra} \and
 \author{W.}{van de Water}}{1997}
            {Effect of depth on the pattern formation of Faraday waves}
            {\PRL}{79}{5010--5013}

\refartbook{refM118}
{\author{H.W.}{M\"{u}ller}, \author{R.}{Friedrich} 
 \and \author{D.}{Papathanassiou}}{1998}
           {Theoretical and experimental investigations of the Faraday 
            instability}
           {Evolution of Spontaneous Structures in Dissipative Continuous 
            Systems}
           {\author{F.H.}{Busse} \and \author{S.C.}{M\"uller}}
           {Springer}{Berlin}{230--265}

\refarticle{refP54}
{\author{E.}{Pamploni}, \author{P.L.}{Ramazza}, \author{S.}{Residori} \and 
\author{F.T.}{Arecchi}}{1995}
            {Two-dimensional crystals and quasicrystals in nonlinear optics}
            {\PRL}{74}{258--261}

\refarticle{refL65}
{\author{S.}{Longhi}}{1999}
            {Transverse patterns in nondegenerate intracavity second-harmonic 
             generation}
            {\PRA}{59}{4021--4040}


\refarticle{refZ9}
{\author{W.}{Zhang} \and {J.}{Vi\~nals}}{1996}
           {Square patterns and quasipatterns in weakly damped Faraday waves}
           {\PRE}{53}{R4283--R4286}


\refarticle{refM126}
{\author{H.W.}{M\"{u}ller}}{1994}
           {Model equations for two-dimensional quasipatterns}
           {\PRE}{49}{1273--1277}  

\refarticle{refL59}
{\author{R.}{Lifshitz} \and \author{D.M.}{Petrich}}{1997}
           {Theoretical model for Faraday waves with multiple-frequency forcing}
           {\PRL}{79}{1261--1264}  

\refarticle{refS105}
{\author{M.}{Silber}, \author{C.M.}{Topaz} \and \author{A.C.}{Skeldon}}{2000}
            {Two-frequency forced Faraday waves:
             weakly damped modes and patterns selection}
            {Physica}{143D}{205--225}

\refarticleshort{refT60}
{\author{C.M.}{Topaz} \and \author{M.}{Silber}}{2002}
            {Resonances and superlattices pattern stabilization in 
             two-frequency forced Faraday waves}
            {Physica, {\rm to appear}}

\refarticleshort{refP53}
{\author{J.}{Porter} \and \author{M.}{Silber}}{2002}
            {Broken symmetries and pattern formation in 
             two-frequency forced Faraday waves}
            {preprint}


\refarticle{refM130}
{\author{B.A.}{Malomed}, \author{A.A.}{Nepomnyashchi\u\i} \and 
 \author{M.I.}{Tribelski\u\i}}{1989}
            {Two-dimensional quasiperiodic structures 
             in nonequilibrium systems}
            {Sov. Phys. JETP}{69}{388--396}

\refarticle{refN18}
{\author{A.C.}{Newell} \and \author{Y.}{Pomeau}}{1993}
            {Turbulent crystals in macroscopic systems}
            {\JPA}{26}{L429--L434}

\refarticle{refS110}
{\author{H.G.}{Solari} \and \author{G.B.}{Mindlin}}{1997}
            {Quasicrystals and strong interactions between square modes}
            {\PRE}{56}{1853--1858}

\refarticle{refE16}
{\author{B.}{Echebarria} \and \author{H.}{Riecke}}{2001}
            {Sideband instabilities and defects of quasipatterns}
            {Physica}{158D}{45--68}


\refarticle{refP56}
{\author{L.M.}{Pismen}}{1981}
            {Bifurcation into wave patterns and turbulence in 
             reaction-diffusion equations}
            {\PRA}{23}{334--344}

\refarticle{refG95}
{\author{A.A.}{Golovin}, \author{A.A.}{Nepomnyashchy} \and 
 \author{L.M.}{Pismen}}{1995}
            {Pattern formation in large-scale Marangoni convection
             with deformable interface}
            {Physica}{81D}{117--147}

\refarticle{refL66}
{\author{P.}{Lyngshansen} \and \author{P.}{Alstr{\o}m}}{1997}
            {Perturbation theory of parametrically driven capillary waves at 
             low viscosity}
            {\JFM}{351}{301--344}

\refarticle{refC120}
{\author{P.}{Chen} \and {J.}{Vi\~nals}}{1999}
           {Amplitude equation and pattern selection in Faraday waves}
           {\PRE}{60}{559--570}

\refarticle{refM25}
{\author{W.V.R.}{Malkus} \and \author{G.}{Veronis}}{1958}
            {Finite amplitude cellular convection}
            {\JFM}{4}{225--260}

\refarticle{refS63}
{\author{A.}{Schl\"uter}, \author{D.}{Lortz} \and \author{F.}{Busse}}{1965}
            {On the stability of steady finite amplitude convection}
            {\JFM}{23}{129--144}

\refbook{refM125}
{\author{J.}{Moser}}{1973}
    {Stable and Random Motions in Dynamical Systems}
    {Princeton University Press}{Princeton}

\refarticle{refI7}
{\author{G.}{Iooss} \and \author{J.}{Los}}{1990}
            {Bifurcation of spatially quasi-periodic solutions in 
             hydrodynamic stability problems}
            {Nonlinearity}{3}{851--871}

\refarticle{refS109}
{\author{J.}{Swift} \and \author{P.C.}{Hohenberg}}{1977}
            {Hydrodynamic fluctuations at the convective instability}
            {\PRA}{15}{319--328}

\refarticle{refM99}
{\author{I.}{Melbourne}}{1999}
            {Steady-state bifurcation with Euclidean symmetry}
            {Trans. Am. Math. Soc.}{351}{1575--1603}

\refbook{refH60}
{\author{G.H.}{Hardy} \and \author{E.M.}{Wright}}{1960}
    {An Introduction to the Theory of Numbers, 4th edition}
    {Clarendon Press}{Oxford}

\refbook{refB126}
{\author{C.M.}{Bender} \and \author{S.A.}{Orszag}}{1978}
    {Advanced Mathematical Methods for Scientists and Engineers}
    {McGraw--Hill}{New York}

\refarticleshort{refD54}
{\author{J.H.P.}{Dawes}, \author{P.C.}{Matthews} \and 
 \author{A,M.}{Rucklidge}}{2002}
            {Reducible actions of $D_4 \sdp T^2$: superlattice patterns and 
             hidden symmetries}
            {preprint}


%
%

\end{thebibliography}
\end{document}